\documentclass[fleqn,usenatbib,twocolumn]{aastex631}
\usepackage{graphicx}
\usepackage{amsmath}
\usepackage{amssymb}
\usepackage{float}
\usepackage{subfigure}
\usepackage{lipsum}
\usepackage[version=4]{mhchem} 
\usepackage{graphicx}
\usepackage{mathtools}
\usepackage{scalerel}
\usepackage{times}
\usepackage{verbatim}
\usepackage{tablefootnote}
\usepackage{savesym}
\savesymbol{tablenum}
\usepackage{longtable}
\usepackage{siunitx}
\restoresymbol{SIX}{tablenum}
\usepackage{multirow}
\usepackage{url}
\usepackage{hhline}
\newif\ifAMStwofonts

\newcommand{\pcm}{\,cm$^{-2}$}	
\newcommand{\rg}{{\thinspace $r_{\rm g}$}}

\newcommand\chidof{{\chi^2/{\rm dof}}}

\newcommand\cdof{{\rm C-stat/{\rm dof}}}

\newcommand\feka{{Fe~K$\alpha$}}
\newcommand\ovii{{O\,\textsc{vii}}}
\newcommand\nvii{{N\,\textsc{vii}}}
\newcommand\neix{{Ne\,\textsc{ix}}}

\newcommand\hbeta{{H$\beta$}}
\newcommand{\nh}{{N_{\rm H}}}
\newcommand\cps{\hbox{$\cts\s^{-1}\,$}}
\newcommand\pscm{\hbox{$\cm^{-2}\,$}}

\newcommand\A{{\rm\thinspace \AA}}
\newcommand\cm{{\rm\thinspace cm}}

\newcommand\erg{{\rm\thinspace erg}}

\newcommand\kev{{\rm keV}}

\newcommand\kpc{{\rm\thinspace kpc}}

\newcommand\m{{\rm\thinspace m}}
\newcommand\pc{{\rm\thinspace pc}}
\newcommand\Msun{\hbox{$\rm\thinspace M_{\odot}$}}

\newcommand\ks{{\rm\thinspace ks}}
\newcommand\s{{\rm\thinspace s}}
\newcommand\cts{{\rm\thinspace count}}
\newcommand\cmps{\hbox{$\cm\s^{-1}\,$}}

\newcommand\ergpscmps{\hbox{$\erg\cm^{-2}\s^{-1}\,$}}

\newcommand\ergps{\hbox{$\erg\s^{-1}\,$}}

\newcommand{\apec}{{\sc apec}}

\newcommand{\chandra}{{\it Chandra}}

\newcommand{\suzaku}{{\it Suzaku}}

\newcommand{\xmm}{{\it XMM-Newton}}

\newcommand{\swift}{{\it Swift}}

\newcommand{\nustar}{\textit{NuSTAR}}

\newcommand{\kms}{km~s$^{-1}$}

\newcommand{\mrk}{Mrk\thinspace1239}
\received{xxx yyy zzz}
\revised{xxx yyy zzz}
\submitjournal{ApJ}
\begin{document}
\title{A Hot Mess: The Rich and Complex Soft Emitting Regions Surrounding the Reflection Dominated Flaring Central Engine of Mrk\,1239}

\correspondingauthor{M. Z. Buhariwalla}
\email{margaret.buhariwalla@smu.com}

\author{M. Z. Buhariwalla}
\affiliation{Department of Astronomy and Physics, Saint Mary's University, 923 Robie Street, Halifax, NS B3H 3C3, Canada}
\author{L. C. Gallo}
\affiliation{Department of Astronomy and Physics, Saint Mary's University, 923 Robie Street, Halifax, NS B3H 3C3, Canada}
\author{J. Mao}
\affiliation{Department of Astronomy, Tsinghua Univerisity, 30 Shuangqing Road, Beĳing 100084, China}
\affiliation{Department of Physics, Hiroshima University, 1-3-1 Kagamiyama, HigashiHiroshima, Hiroshima 739-8526, Japan}
\author{J. Jiang}
\affiliation{Institute of Astronomy,  Madingley Road, Cambridge CB3 0HA, UK}
\author{L. K. Pothier-Bogoslowski}
\affiliation{Department of Astronomy and Physics, Saint Mary's University, 923 Robie Street, Halifax, NS B3H 3C3, Canada}
\author{E. J{\"a}rvel{\"a}}
\affiliation{Homer L. Dodge Department of Physics and Astronomy, The University of Oklahoma, 440 W. Brooks St., Norman, OK 73019, USA}
\author{S. Komossa}
\affiliation{Max-Planck-Institut f{\"u}r Radioastronomie, Auf dem H{\"u}gel 69, 53121, Bonn Germany}
\author{D. Grupe}
\affiliation{Department of Physics, Geology, and Engineering Technology Science Center, Northern Kentucky University, 1 Nunn Drive, Highland Heights, KY 41099}

\shorttitle{A Hot Mess: Mrk 1239}
\shortauthors{Buhariwalla et al. 2024}
\begin{abstract}
Previous X-ray works on \mrk\ have revealed a complex Narrow Line Seyfert 1 (NLS1) that exhibits substantial absorption and strong emission from both collisional (CIE) and photoionized (PIE) plasmas. Here, we report on deep-pointed observations with \xmm\ and \nustar, along with \swift\ monitoring, to understand the $0.3-30$  \kev\ continuum emission and the central engine geometry. A strong X-ray flare, where the AGN brightens by a factor of five in $\sim30$\ks, is captured between $4-30$  \kev\ and can be attributed to a brightening of the primary continuum. However, the lack of any variability below $\sim3$  \kev\ on long- or short-time scales requires complete absorption of the AGN continuum with a neutral medium of column density $\sim 10^{23.5}$\pscm. The timing and spectral properties are consistent with a blurred reflection interpretation for the primary emission. The variability and presence of a Compton hump disfavours ionized partial covering. The neutral absorber, if outflowing, could be crashing into the surrounding medium and ISM to produce the low-energy continuum and CIE.  Scattered emission off the inner torus could produce the PIE. The intricate scenario is demanded by the data and highlights the complexity of the environment that is normally invisible when overwhelmed by the AGN continuum. Objects like \mrk\ serve as important sources for unveiling the interface between the AGN and host galaxy environments.
\end{abstract}



\section{Introduction}

According to the Unified Model (e.g. \citealt{Antonucci+1993}, \citealt{Urry+1995}, \citealt{Urry+2003}), all active galactic nuclei (AGN) are fundamentally the same. They consist of a supermassive black hole (SMBH) surrounded by an accretion disc \citep[e.g.][]{Shakura+1973}, a broad line region (BLR) \citep[e.g.][]{Peterson+1999, Peterson+2004}, and a dusty torus \citep[e.g.][]{Netzer+2013, Peterson+1997}. These components are thought to exist in all AGN, and the various AGN classifications indicate different viewing angles through the obscuring torus. Type I AGN offer a clear, unobstructed view of the accretion disc and BLR with low inclinations and low column densities. In contrast, Type II AGNs are thought to have much more edge on view through the obscuring torus \citep{Antonucci+1993}, effectively absorbing all the soft (less than 3  \kev) photons emitted from the AGN. 

In AGN, the primary X-ray source is considered to be the corona, a cloud of hot electrons located above the accretion disc that Compton-up scatter UV disc photons to higher energies \citep{Haardt+1993}. Some of these photons are emitted directly as the primary X-ray continuum, while others are reflected by the accretion disc and surrounding material, producing reflection spectra \citep[e.g.][]{George+1991}. Photons reflected near the SMBH produce a blurred ionized spectrum due to the immense relativistic effects \citep[e.g.][]{Tanaka+1995, Laor+1991, Ballantyne+2001, RossFabian+2005}, while photons reflected at larger distances produce narrow, near neutral emission features \citep{Nandra+1997}. The hard X-ray spectra above 10  \kev\ can be attributed to the primary continuum, Compton hump, synchrotron self-Comptonization (SSC) from radio jets \citep{Netzer+2013} or any combination of these components. 

AGN can shape their host galaxy by way of kinetic or radiative feedback \citep[][and ref. therein]{Morganti+2017}. We see evidence of this  in the M-$\sigma$ relation, where the mass of the SMBH scales with the stellar velocity dispersion of the halo stars \citep[e.g.][]{Gultekin+2009}. Kinetic feedback is driven by radio jets shooting out of the central engine and interacting with the host galaxy  \citep[e.g.][]{Fabian+2012B}.
Radiative feedback is dominant when young nuclei are surrounded by cold neutral gas that can be radiatively driven away from the central engine \citep[e.g.][]{Fabian+2012B}. Depending on the Eddington ratio of the source and the column density of the obscuring cloud, clouds can be either long-lived and stable or outflowing   \citep[e.g.][]{Fabian+2009A, Ishibashi+2018}. 

Warm absorbers are a common absorption component found in Seyfert 1 galaxies \citep[e.g.][]{Crenshaw+2003}. They consist of partially ionized outflowing clouds located close to the central engine of the AGN. They typically have column densities less than $\nh=10^{23}\cmps$, are outflowing at $\sim1000$ \kms\ and contain little dust.

A very interesting subclass of AGNs are the Narrow Line Seyfert 1s (NLS1s), which are characterized by the widths of permitted optical lines originating from their BLR. This classification was introduced by \cite{Osterbrock+1985}, and it is based on the width of the broad \hbeta\ line. Seyfert 1s that have \hbeta\ widths less than 2000 \kms\ are called NLS1; if the \hbeta\ widths is greater than 2000 \kms\, then they are called Broad Line Seyfert 1 (BLS1). NLS1s are thought to be the younger counterparts to BLS1, with lower black hole masses \citep{Grupe+1996, Mathur+2000}. They presumably accrete at high Eddington fractions \citep{Pounds+1995, Grupe+2004B, Komossa+2008}, and are typically hosted in spiral galaxies \citep{Seyfert+1943, DeRobertis+1998, Varglund+2022}. While NLS1s can host radio jets they are generally radio-quiet objects \citep{Komossa+2006, Foschini+2012, Berton+2015, Jarvela+2022}. 

The X-ray spectrum of NLS1s is often very variable both in flux and spectral shape on rapid and long-term time scales \citep{Leighly+1999A, Leighly+1999B, Gallo+2018}. The rapid variability seen in these objects indicates a physically small emitting region \citep{Ponti+2012, Fabian+2015}. They typically have steep photon indices \citep{Brandt+1997, Leighly+1999B} indicating cooler and/or more diffuse corona \citep{Fabian+2015}. 

NLS1s often show an excess in the soft X-ray band ($<3$  \kev) \citep{Puchnarewicz+1992, Boller+1996, Grupe+1998, Waddell+2020}, the origin of which is still up for debate. This spectral feature is present in BLS1s but is stronger in NLS1s \citep[e.g.][]{Waddell+2020}. It could be the result of partial covering of the central engine \citep[e.g.][]{Tanaka+2003, Tanaka+2004, Gierlinski+2004}, or the result of relativistically blurred soft emission lines from the accretion disc \citep{Ballantyne+2001, Fabian+2009B}. The soft excess may also originate from a secondary warm corona existing on top of the accretion disc \citep[e.g.][]{Magdziarz+1998}. The soft excess may have different mechanisms depending on the source. Typically, it is one of the most variable components of the X-ray spectra of an AGN \citep[e.g.][]{Boller+1996, Leighly+1999B}. 

\mrk\ may be unlike any other NLS1 studied to date. It presents remarkable challenges and unique features in all energy bands that it has been observed in. \mrk\ exhibits bipolar radio jets on both parsec (pc) and kiloparsec (kpc) scales \citep{Doi+2013, Doi+2015}. They appear to be roughly symmetric, subsonic, one approaching and one receding, though which one is which is undetermined. The opening angle also cannot be determined because the location of the radio core is unknown \citep{Doi+2015}. 

\mrk\ straddles the border between radio-quite \citep{Doi+2015} and radio-loud \citep{Berton+2018} and is classified as a Fanaroff-Riley type I (FRI) radio source \citep{Fanaroff+1974, Doi+2015}, where radio emission is most intense in the center of the galaxy and decreases intensity with increasing radius. \cite{Buhariwalla+2023} overplotted the soft band emission and the radio contours produced by \cite{Jarvela+2022}  to show that the soft X-ray emission is asymmetric and extended from the center of the galaxy (see Fig 1 of \citealt{Buhariwalla+2023}). 

\mrk\ shows an intriguing infrared spectrum, with a blackbody emission feature attributed to hot dust from the torus \citep{Rod+2006}, with a dust temperature of  $T=1210 \,{\rm K}$. Polycyclic aromatic hydrocarbon (PAH) emission has been detected in the central $\sim400\,{\rm pc}$ of \mrk\ indicating an upper limit of star formation of 7.5 \Msun yr$^{-1}$ \citep{RuschelDutra+2017}. 

The optical spectra of \mrk\ were used in the original definition of NLS1s by \cite{Osterbrock+1985}. The FWHM of the \hbeta\ has been recently measured to be between 800\,\kms\ \citep{Gravity+2023} and 1300\,\kms\ \citep{Husemann+2022}. The  extinction of \mrk\ is measured to be $E_{B-V} = 1.6$ \citep{Pan+2021,Feltre+2023}. \mrk\ is a highly polarized source \citep{Goodrich+1989, Pan+2019, Pan+2021} whose optical band shows evidence of two distinct regions of polarization. One of which is thought to be associated with the hot dust seen in the NIR band \citep{Rod+2006}. 

The X-ray band is where \mrk\ is truly unique. The broadband continuum of this object shows an incredibly strong soft excess, a possible broad iron line and a Compton hump \citep{Buhariwalla+2020}. The continuum was originally reported to be heavily obscured by a partial covering absorber with a column density in excess of $\nh=3\times10^{23}$\pcm\  \citep{Grupe+2004}. 

Excess emission at 0.9  \kev\ was originally reported as the \neix\ triplet \citep{Grupe+2004}. Due to the absence of the \ovii\ triplet, the source was inferred to have an overabundance of Ne. Later, this feature was interpreted by \cite{Buhariwalla+2020} as collisionally ionized emission (CIE), possibly from a region of star formation in the host galaxy. The star formation rate (SFR) measured using the X-ray spectra was $\simeq4-6$\Msun\,yr$^{-1}$, while the SFR predicted in other bandpasses is  $\leq$ 7.5\Msun\,yr$^{-1}$ from PAH measurements \citep{RuschelDutra+2017}; $3.47\pm 0.26$\Msun\,yr$^{-1}$ from SED fitting \citep{Gruppioni+2016}; and  $2.1_{-0.4}^{+0.5}$\Msun\,yr$^{-1}$ from IR measurement \citep{Smirnova+2022}. 

\cite{Buhariwalla+2020} performed a multi-epoch, broadband analysis of all X-ray data available to attempt to understand this galaxy. The soft excess of \mrk\ is observed to be remarkably constant over a $\sim20$ year period. This was attributed to the physically large scale of the CIE visible because of the heavy absorption of the AGN continuum. The AGN continuum was again interpreted as partial covering; however, blurred reflection was not ruled out. Further data were needed to explore the soft band emission and AGN continuum. 

The first high-resolution spectrum of \mrk\ was reported by \cite{Buhariwalla+2023}. Ionized emission lines were detected between $0.4-1.7$  \kev\ ($7-35$ \AA); however, due to the dim nature of the source, no continuum was observed. The ionized emission was determined to originate from two distinct plasmas: the CIE and a second photoionized plasma (PIE). The photoionized plasma was found to be ionized by the AGN at a torus-like distance and possibly outflowing at $\sim660$\,\kms. The SFR measured was $\sim3$\Msun\,yr$^{-1}$, consistent with other measurements. The presumed large scales of the plasma were consistent with the extended and asymmetric X-ray emission seen in the \chandra\ image \citep{Buhariwalla+2023}. 

This paper presents an X-ray broadband timing and spectral analysis of \mrk\ with new data from \xmm, \nustar, and \swift. These data were collected primarily in late 2021 and include a joint \xmm\ and \nustar\ observation with two \swift\ monitoring campaigns. In Section \ref{sec:data}, the new observations and data reduction techniques are summarised. Section \ref{sec:TIME} examines the timing analysis, including the rapid and long-term variability. Light curves, hardness-flux diagrams and principal component analysis are explored. In Section \ref{sec:spec}, the spectra are analyzed, beginning first with the soft band below 2  \kev, then the hard band above 4  \kev, and finally, the broadband $0.3-25$  \kev\ spectra are analyzed. A spectral variability analysis is completed in Section \ref{sec:spec_var}. The results are discussed in Section \ref{sec:disc}, and conclusions are drawn in Section \ref{sec:conclusion}. 

\section{Observations and Data Reduction}
\label{sec:data}
\begin{table*}
\centering
	\begin{tabular}{c c c c c c c c c}
	\hline
	(1)         & (2)            & (3)             & (4)    & (5)        & (6)     & (7)      & (8)    &(9)              \\
	Observatory & Observation ID & Instrument Name & Label  & Start date & Duration& Exposure & Counts & Energy range	 \\
                    & 	             &       		   &        &(yyyy-mm-dd)& [s]     & [s]      &    	   & 			 	 \\
	\hline
	\xmm		& 0891070101	 & PN 			   &\multirow{2}{*}{Low State}    &2021-11-04  & 105000  & 79175	  & 42042  & $0.3-10$ \kev   \\
	\nustar     & 60701038002	 & FPMA/FPMB 	   &   & 2021-11-04 & 110050  & 101641   & 8841   & $4-30$  
 \kev$^*$\\ 
  \hline
                    &                & FPMA/FPMB 	   & High State & 2021-11-05 & 106689  & 99177	  & 17125  & $4-30$  \kev$^*$\\ 
	\hline
        \swift      & 00031685,00081986 & XRT          &SwiftAve&   -        & -       & 39777    &  1094  &$0.3-10$  \kev   \\
    \hline	
	\end{tabular}
	\caption{Observations log for Mrk 1239. The observatory used for analysis is listed in Column (1). The observation ID and instrument name are given in columns (2) and (3), respectively. The label used in this work is given in Column (4). The start date of each observation is given in Column (5). The duration of each observation, total exposure time and total counts for each observation are given in columns (6), (7), and  (8), respectively. The energy of each observation used is given in Column (9). The summed duration, exposure and counts are reported for \nustar\ FPMA and FPMB. $^*$This is the energy range used for all \nustar\ light curves unless otherwise stated. The energy range used for spectral fitting was $4-25$\kev\ due to loss of signal to notice at higher energies in the uncombined spectra.}
	\label{tab:obs}
\end{table*}
Observations of \mrk\ began on 2021-11-04 with both \xmm\ \citep{Jansen+2001} and \nustar\ \citep{nustar+2013}. The \xmm\ observation concluded on 2021-11-05, while the \nustar\ observation concluded on 2021-11-06. Each observatory obtained $\sim100$ \ks\ of on-source time. \mrk\ entered a flaring state soon after the \xmm\ observation concluded (see Figure \ref{fig:LIGHTCURVE}, and Section \ref{sec:nustar} for details). The data used for spectral analysis is listed in Table \ref{tab:obs}. A second data set comprising \swift\ \citep{Burrows+2004} snapshots was also obtained and will be present here; the data for these observations is listed in Table \ref{tab:SwiftObs}.

\subsection{\xmm}
The \xmm\ data were processed using the \xmm\ Science Analysis System, {\sc sas} V17.0.0. An event list was created from the Observation Data Files (ODF) using {\sc epproc} and {\sc emproc} for the PN and MOS instruments, respectively. Light curves were generated using {\sc evselect}, and {\sc tabgtigen} was implemented to remove all times when the background had evidence of flaring. No evidence of pileup was found in the PN and MOS data. For both the PN and the MOS spectra, the source data was extracted from a circular region 35"  centred on \mrk, while the background was extracted using a circular region of 50" on the same chip. Response files were generated using {\sc rmfgen} and {\sc arfgen}. We checked whether changes of the PN effective area measured empirically with \nustar, which were outlined in Technical Note 230\footnote{ \url{www.xmmweb.esac.esa.int/docs/documents/CAL-TN-0230-1-3.pdf}} (hear after TN230) had a significant impact on our conclusions. This was done by running {\sc arfgen} with the flag {\sc applyabsfluxcorr=yes} included. We find that including this flag had a minimal effect on PN spectra. There was not a significant impact on our measured parameters or conclusions. Only the PN data were used for spectral modelling, while the MOS data were used for qualitative comparisons and consistency checks. 
\subsection{\nustar}

The FPMA and FPMB data were extracted from a source region of 75". A background was selected from the same chip with a region of 120". The data were processed with {\sc caldb} index version 20210315, using {\sc nuproducts}. Good time intervals (GTI) were generated using {\sc gtibuild} and used to extract a low flux (low state) and high flux (high state) spectrum for both FPMA and FPMB (see Section \ref{sec:nustar} for details). The \nustar\ data were optimally binned using {\sc ftgrouppha}. 

We note a difference in flux between the \nustar\ and PN flux between $3-4$ \kev. We determined it was not due to a tear in the Multi-Layer Insulation (MLI), as the discrepancy was seen in both detectors, not just FPMA \citep{Madsen+2020}. We also tested to see if the discrepancy was due to changes in the effective area of PN as per TN230; again, we found this correction insufficient for the data. We note the source is relatively dim at 3 \kev, approaching the background level in \nustar. Thus, ignoring all \nustar\ data below 4 \kev\ does not influence the results.

\subsection{\swift} 
\begin{table}
\centering
	\begin{tabular}{c c c c }
		\hline
    (1) 		  & (2) 		 &(3) 		    & (4)          \\
Observation ID    & Start date   & Duration 	& Count Rate	      \\
		         & (yyyy-mm-dd) & [s]		   & [cps]	\\
	   \hline
    00031685001   & 2010-04-13     &   3616      & $0.035\pm0.004$\\
\textcolor{white}{.} 00081986001$^{\dagger}$   &	2019-06-17    &   6216 & $0.047\pm0.003$ \\
   \hline
    00031685002   &   2021-10-29  &   1596		&	$0.047\pm0.006$\\
    00031685003	  &   2021-10-31  &   1431 		&	$0.030\pm0.005$\\
    00031685004	  &   2021-11-02  &   1573		&	$0.034\pm0.005$\\
    00031685005   &   2021-11-04  &   1648 		&	$0.054\pm0.007$\\
    00031685006	  &   2021-11-06  &   1491 		&	$0.062\pm0.008$\\
    00031685007	  &   2021-11-08  &   1618  	&	$0.059\pm0.007$\\
    00031685008	  &   2021-11-10  &   814 	 	&	$0.038\pm0.008$\\
    00031685009	  &   2021-11-12  &   1683 		&	$0.083\pm0.008$\\
    00031685010	  &   2021-11-14  &   1701 		&	$0.039\pm0.008$\\
    \hline
    00031685011   &   2022-11-23  &   1988      &   $0.031\pm0.005$\\
    00031685012   &   2022-11-26  &   1532      &   $0.032\pm0.005$\\
    00031685013   &   2022-11-29  &   2053      &   $0.031\pm0.005$\\
    00031685015   &   2022-12-05  &   2058      &   $0.019\pm0.004$\\
    00031685016   &   2022-12-08  &   1965      &   $0.021\pm0.004$\\
    00031685017   &   2022-12-11  &   1439      &   $0.034\pm0.006$\\
    00031685018   &   2022-12-15  &   1231      &   $0.023\pm0.006$\\
    00031685019   &   2022-12-18  &   2151      &   $0.025\pm0.004$\\
    00031685020   &   2022-12-21  &   2087      &   $0.050\pm0.011$\\
  \hline	
	\end{tabular}
	\caption{Observations log for the \swift\ monitoring campaign of Mrk 1239. The \swift\ observation IDs are listed in Column (1). The start date of each observation is given in Column (2),  and the duration of each observation and the $0.3-10$  \kev\ count rates for each observation are given in Columns (3) and (4), respectively. $^{\dagger}$This observation was simultaneous with the 2019 \nustar\ observation. The table is divided into three sections: the first contains two historical observations,  the second is the \swift\ monitoring surrounding the deep \xmm-\nustar\ observation, and the third is the second monitoring campaign.}
	\label{tab:SwiftObs}
\end{table}

In the weeks surrounding the deep observations of \mrk, \swift-XRT took several snapshot observations of the AGN approximately every two days. In December 2022, ten more snapshot observations were taken. These observations, along with the two archival observations, were compiled for analysis. The details of these observations can be found in Table \ref{tab:SwiftObs}.

The  \swift\ XRT spectra were created from the \swift-XRT data product generator (\citealt{Evans+2009})\footnote{\url{www.swift.ac.uk/user_objects/}}. A spectrum was extracted for each observation ID listed in Table \ref{tab:SwiftObs}. The data were binned into ten bins using {\sc grppha}. Individual \swift\ spectra can be seen in  Appendix \ref{A:swift}. The light curve binned by observation ID was also extracted using the product generator with two energy bins, $0.3-3$  \kev\ and $3-10$  \kev.

A final average \swift\ spectrum was also extracted using \swift-XRT data product generator. It contains all the data in Table \ref{tab:SwiftObs}, resulting in a total exposure time  $\sim40$\ks. The spectrum was optimally binned using {\sc ftgrouppha}.  

\section{Timing Analysis}
\label{sec:TIME}
\cite{Buhariwalla+2020} described the soft band ($<2$  \kev) of \mrk\  with three distinct components; the first was the collisionally ionized component, which was interpreted as originating from a region of star formation in the host galaxy.  The second was a photoionized component, interpreted as reflection off a torus-like structure. The final component was a relatively featureless component, interpreted as the AGN continuum leaking through the obscuring clouds. The components and interpretation were largely consistent in the RGS analysis of \cite{Buhariwalla+2023}. In the RGS spectra, the photo and collisionally ionized components are visible, while the third continuum-like component was not distinguished from the background. 

One of the driving characteristics that led to the development of the three-component soft band was the unique variability behaviour of \mrk. This object has shown remarkable consistency in the soft band (below 3  \kev) while exhibiting typical levels of AGN variability in the hard band (above 3 \kev). These new data demonstrate that behaviour to an extreme, placing more restrictions on the primary absorber that obscures the central engine in \mrk\ and the nature of the featureless component. In this section, we probe time scales ranging from years long with the unfolded spectrum (Sec. \ref{sec:Long}) to month-long monitoring with two \swift\ monitoring campaigns (Sec. \ref{sec:Inter}); to hours with the PN and \nustar\ light curves (Sec. \ref{sec:nustar}). This allows us to probe the variability of \mrk\ like never before. We will conduct a Principal Component Analysis (PCA) in Section \ref{sec:PCA}. Finally, we will conclude what the timing analysis of \mrk\ can tell us about this unique source in Section \ref{sec:TIMEconc}.

\subsection{Long-term Time Scales}
\label{sec:Long}
\begin{figure}
\centering
\includegraphics[width=\linewidth]{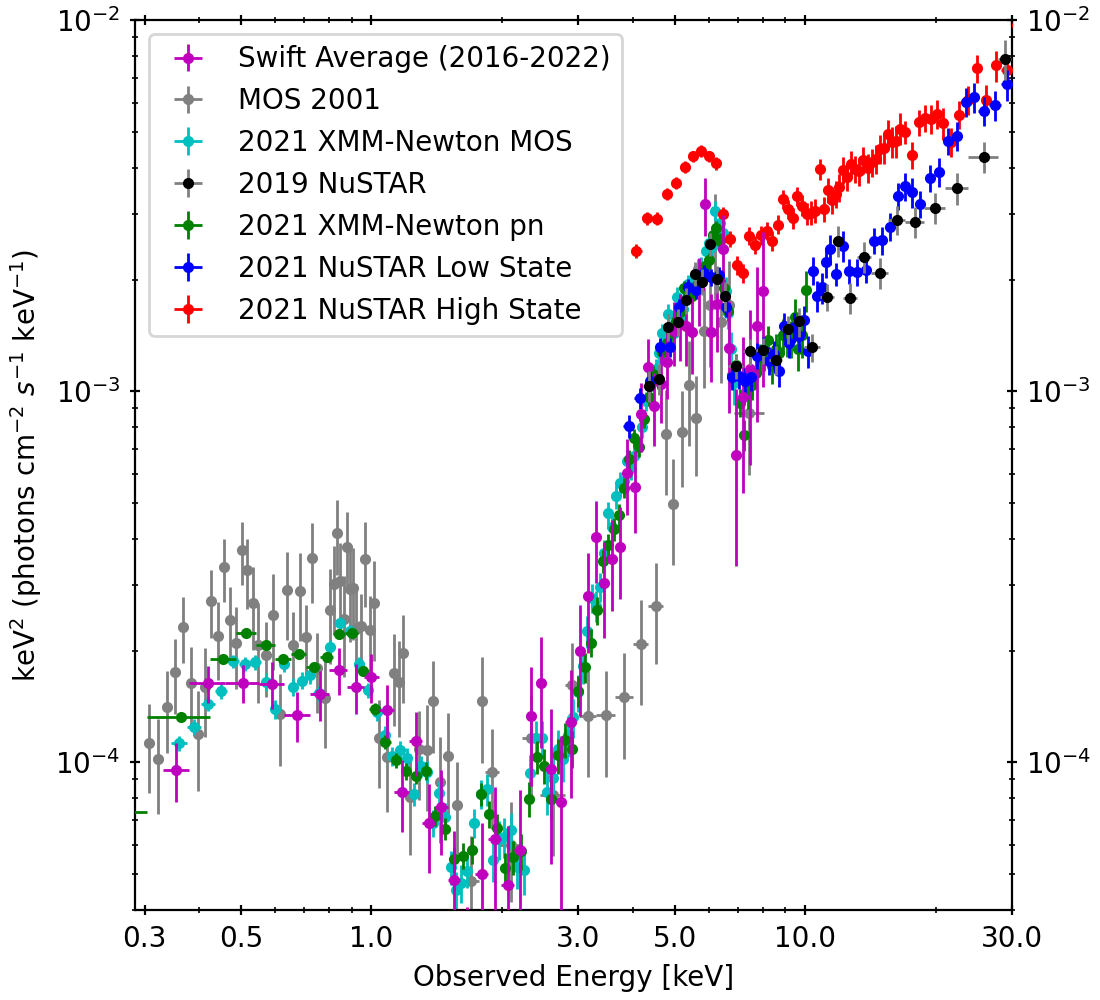}
    \caption{The unfolded spectra of \mrk, showcasing the new pointed 2021 observations with PN (green), \nustar-low state (blue) and \nustar-high state (red). For comparison, the average \swift\ spectra (magenta), the 2001 MOS spectra (gray) and 2019 \nustar\ (black)  are also shown.}
    \label{fig:EEUF}
\end{figure}
We begin on the longest time scales, comparing data spanning 20 years. The unfolded spectra of \mrk\ compared to a flat ($\Gamma=0$) power law are presented in Figure \ref{fig:EEUF}. The 2001 MOS, 2019 \nustar, and an average \swift\ spectrum are included to compare how the general spectral shape has changed in the 20 years of observations. 

In the soft band below 3 \kev, the PN and average \swift\ spectra appear consistent, and both spectra generally agree with the MOS spectra. There has been very little change in the flux and shape of the spectra over the 20 years of observation. This is consistent with the results of \cite{Buhariwalla+2020}. Differences appear between 3 and 5  \kev, where the more recent data is at a higher flux. This does not appear to be a calibration effect arising from the off-axis MOS observation. It may be a decrease in the column density of the neutral absorber that is covering the primary continuum. This is consistent with the previous spectral fitting results in \cite{Buhariwalla+2020}.
\begin{figure*}
\centering
\includegraphics[width=\linewidth]{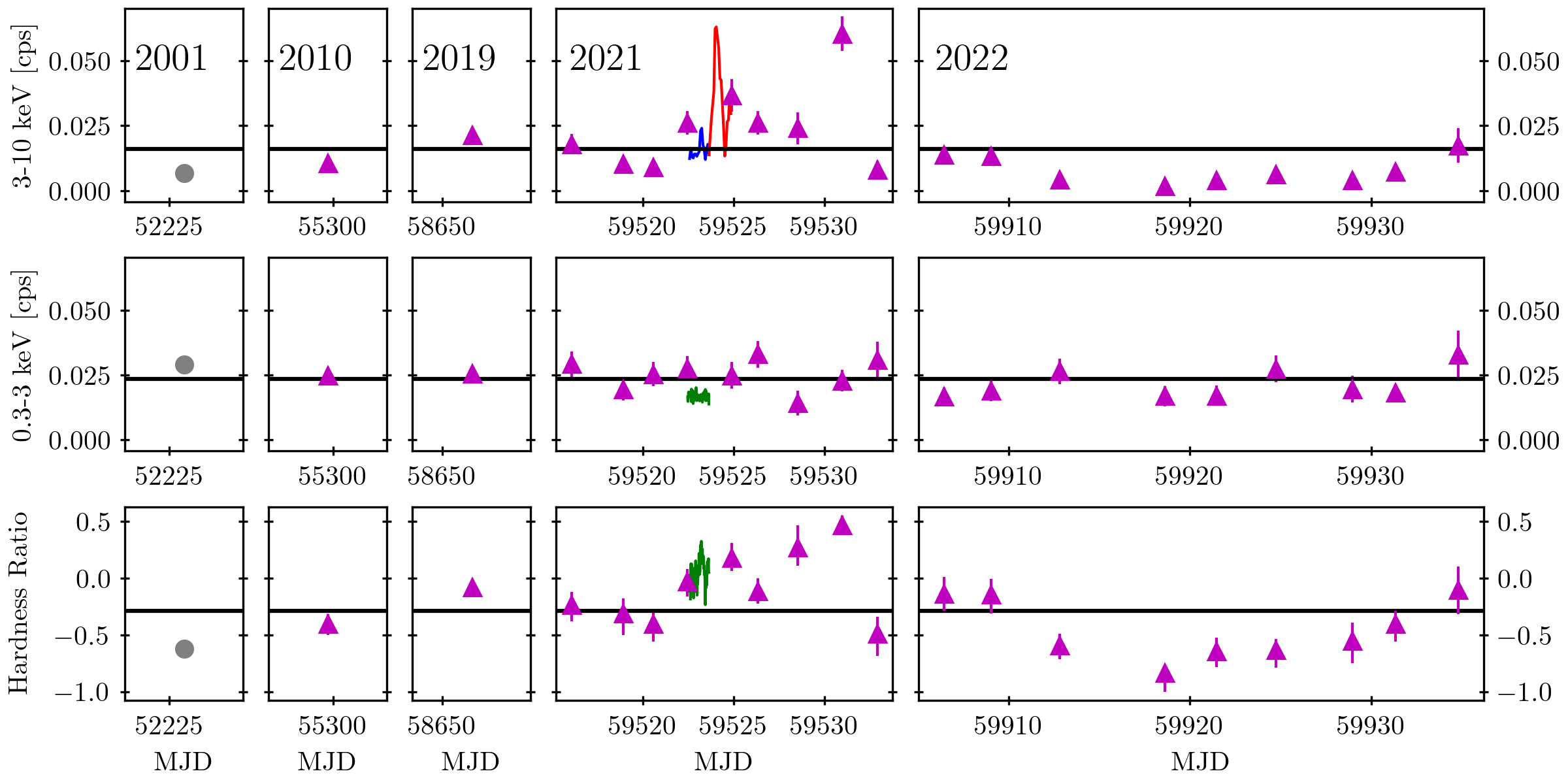}
    \caption{The light curves of \mrk\ for the \swift-XRT monitoring campaign, with MOS and 2021 \nustar\ data projected into \swift\ count rates (see text for details). Each column presents data from a different epoch and the year of the epoch can be found in the top panel.  \textit{Top panel:} The hard band ($3-10$  \kev) light curve for XRT is shown in magenta, while the $3-10$  \kev\ light curve for \nustar\ is shown in red (high state) and blue (low state). The 2001 MOS data is shown in the first column and is in grey.  \textit{Middle panel:} The soft band ($0.3-3$  \kev) for XRT is magenta, and the average \swift\ count rate is drawn in black. Notice the remarkable constancy displayed in the soft band. For completeness, the PN $0.3-3$ \kev\ light curve is also projected into a \swift\ count rate. It is shown in green. \textit{Bottom panel:} The normalized hardness ratio of the XRT light curves, with the soft and hard bands shown in the panels above. The  \swift\ average hardness ratio is shown in the black line. For completeness the PN hardness ratio is shown in green, calculated in the same bands that the \swift\ hardness ratio is.}
    \label{fig:jun19_swift_lc}
\end{figure*}

A flare was captured during the 2021 \nustar\ observation (see Section \ref{sec:nustar} for details). To compare the spectral variability during the flaring periods, the \nustar\ data was divided into a low state and a high state. Figure \ref{fig:LIGHTCURVE} shows this divide.  The low state \nustar\ spectrum is simultaneous with the PN spectra and appears nearly identical to the 2019 \nustar\ spectrum. The low state spectrum appears slightly brighter above 20  \kev. The high state spectrum is brighter than the low state spectra until approximately 20  \kev. The rapid variability between $10-20$  \kev\ indicates continuum changes in the central engine during the flare. 

The long-term light curves for this source are presented in Figure \ref{fig:jun19_swift_lc}. Here, we show 20 \swift\ observations spanning 12 years; the spectra for these data can be seen in Appendix  \ref{A:swift}. We also include  \swift-projected \nustar\ and \xmm\ light curves for the 2021 \nustar\ and PN data along with the 2001 MOS data. To project the \nustar\ light curves into \swift-XRT count rates, a conversion factor was estimated using WebPIMMS\footnote{\url{https://heasarc.gsfc.nasa.gov/cgi-bin/Tools/w3pimms/w3pimms.pl}}, and the average \nustar\ spectra fit with an absorbed power law over $4-10$  \kev. 

Similarly, the MOS and PN spectra were fitted with an empirical model ({\sc bb+po}) to obtain a conversion to the \swift\ count rate using WebPIMMS. The  $3-10$  \kev\ estimated PN count rate was indistinguishable from the \nustar\ low state light curve and thus has been left off the figure.  The  $0.3-3$  \kev\ light curve and hardness ratio have been included in Figure \ref{fig:jun19_swift_lc}.

Throughout this work, we use the normalized hardness ratio given by $HR = \frac{H-S}{H+S}$, where $H$ and $S$ are the hard and soft bands, respectively. The exact hard and soft bands used will be clearly stated. The data has sufficient counts in the \xmm\ and \nustar\ light curves to be treated as Gaussian. However, the \swift\ snapshots contain limited counts and thus fall into the Poisson regime. For these data, we utilize {\sc behr}\footnote{Bayesian Estimation of Hardness Ratios \url{https://hea-www.harvard.edu/AstroStat/BEHR/}} \citep{Park+2006}. The value of the hardness ratio is taken from the median, and the uncertainties are at 1$\sigma$.

We also use fractional variability ($F_{var}$) to explore the timing properties of the \swift\ data and utilized the method given by  \cite{Edelson+2002}\footnote{See appendix A1 of \cite{Edelson+2002} for the derivation of the uncertainty.} to calculate the error bars ($\sigma_{F{\rm var}}$), specifically:
\begin{equation}
F_{{\rm var}} = \sqrt{\frac{S^2-\langle \sigma^2_{{\rm err}}\rangle}{\langle X\rangle^2}}, \quad 
\sigma_{F{\rm var}} = \frac{1}{F_{{\rm var}}} \sqrt{\frac{1}{2N}} \frac{S^2} {\langle X \rangle^2}\,.
\label{eq:fvar}
\end{equation}
Here $S^2$ represents the variance of the data, $\langle X \rangle$ is the mean count rate,  $\langle \sigma^2_{{\rm err}}\rangle$ is the mean of the square of the standard error, and $N$ is the number of data points in the light curve. 

The long-term \swift\ light curve shows the extreme variability of \mrk\ above 3 \kev. Between 2001-2022, the hard band exhibits fluctuations around the average hard count rate. While the soft count rate remains relatively consistent. The fractional variability of the \swift\ soft data is $12\pm8$\%, while the fractional variability of the hard band is $83\pm14$\%. We caution, however, as the Edelson method assumes Gaussian data and the \swift\ hard band is in the Poisson regime.  Despite this, there appears to be a stark contrast between the variability in amplitude in each band, indicating a disconnect between the hard and soft emitting regions. 

Comparing the \swift\ soft band to its average value results in $\chidof=28/19$ ($p=0.08$, where $p$ is the probability of the null hypothesis that the light curve is constant in time). In contrast, compared to their mean values, the hard band and the hardness ratio highlight significant variability, resulting in a fit statistic of $\chidof=353/19$ ($p<0.0001$) and $\chidof=152/19$ ($p<0.0001$), respectively.  While we can neither accept nor reject the null hypothesis that the data is constant in time for the soft band, we can reject the null for the hard band and the hardness ratio.   It appears there is a total disconnect between the soft and hard bands in \mrk.

\subsection{Intermediate Time Scales}
\label{sec:Inter}
Looking at each of the approximately month-long \swift\ light curves in isolation, we can begin to probe the variability of this source on intermediate timescales. During the 2021 monitoring campaign, there are rapid changes in the hard flux, suggesting some flaring activity.  One of these flaring events occurred during the \nustar\ observation. Another occurred several days later on November $12^{th}$ (MJD 59530), suggesting that \mrk\ was undergoing multiple flaring events during this epoch.  The  \swift\ soft band shows no evidence of change in the count rate during or after this observation. 

In 2022, the hard count rate drops to as low as 10\%\ the average count rate. These low-flux events appear with a drastic softening of the spectra.  It is similar to the count rate and hardness ratio observed in \mrk\ during the 2001 \xmm\ observation, while the source was obscured with a higher column density neutral absorber.  A second average \swift\ spectrum was created using the softest \swift\ observations from 2022. The spectrum is comparable to the 2001 MOS spectrum above 3  \kev, suggesting that the source was in a similar state at the time of these observations.  All other observations of \mrk\ since 2019 have shown a less absorbed hard continuum compared to  2021.  Examining the general shape of the 2022 hardness ratio where the data starts near the average, then drops to the lowest hardness ratio observed by \swift\ before returning to the average almost a month later, a possible explanation could be a transiting obscuration event such as the one seen in NGC 6814 \citep{Gallo+2021} but on much longer time scales.  More data would be required to confirm such transient events. 

Based on the \swift\ monitoring and data above 3 \kev, the continuum of \mrk\ undergoes periods of rapid flux changes and intense dimming. The nature of the continuum variability will be investigated below.
\subsection{Rapid Time Scales}
\label{sec:nustar}
\begin{figure*}
	\centering
\includegraphics[width=\linewidth]{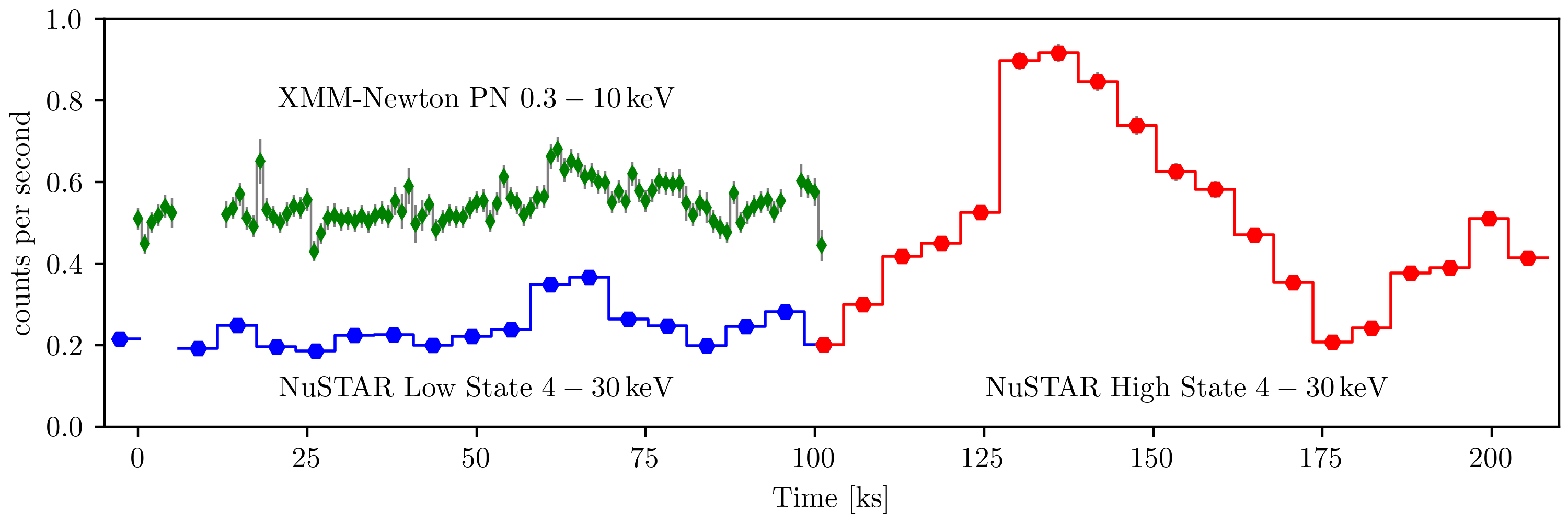}
	\caption{Light curves from the deep \xmm\ and \nustar\ observations. A flaring event was detected during the later half of the \nustar\ observation.  The data were divided into low and high states. This divide illustrates the times when the \nustar\ spectra are extracted in low and high states. The  PN data are in green diamonds, while the NuLow and NuHigh data are in blue and red hexagons, respectively.  For all \nustar\ data presented, FPMA and FPMB are summed. Zero seconds marks the start of the PN exposure.}
	\label{fig:LIGHTCURVE}
\end{figure*}

\begin{figure}
\centering
\includegraphics[width=\linewidth]{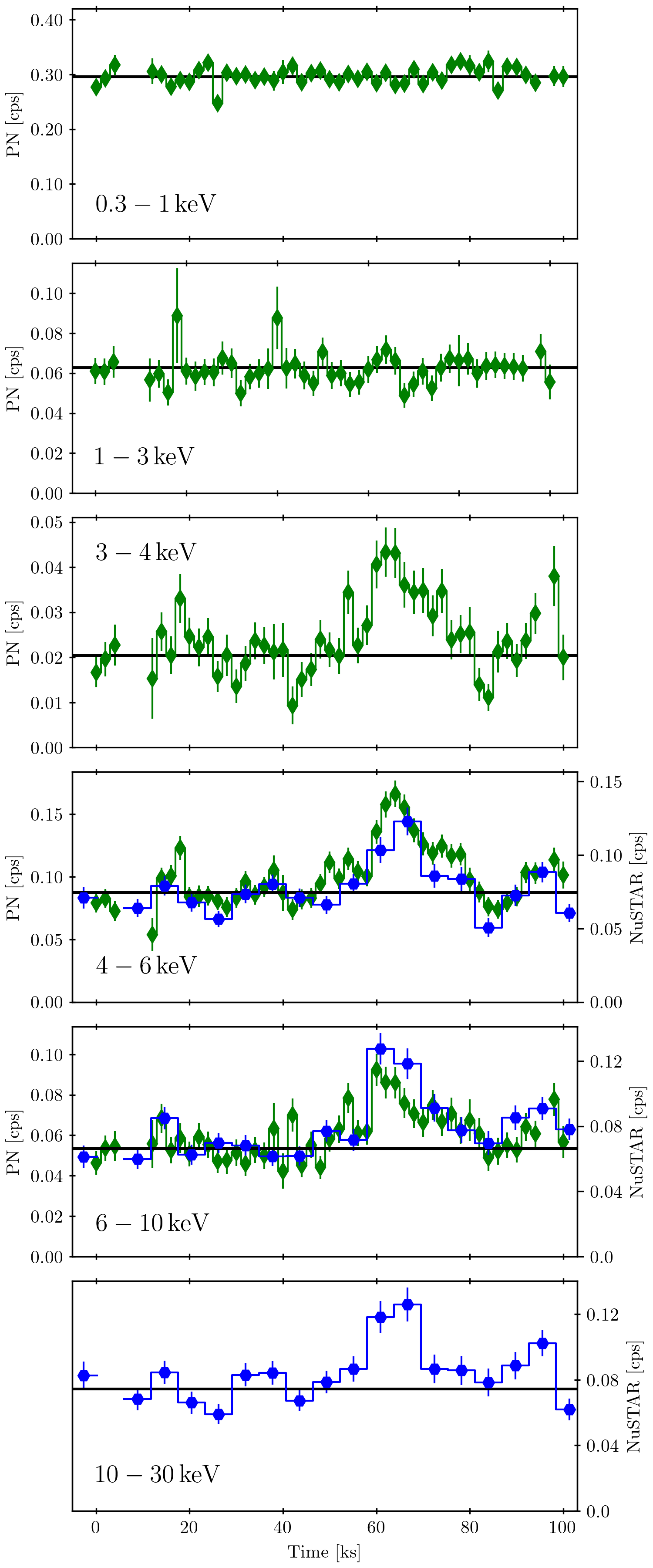}
    \caption{Low state light curves for the PN (green diamonds; left axis) and \nustar\ (blue hexagons; right axis) data. The PN light curves are binned to 2000\,s and divided into five energy bins. The  \nustar\ light curves are binned into orbital bins (5780\,s) and divided into three energy bins. The energy bins are listed in the panel. Where the instruments overlap in energy, both light curves are shown.  The black line shows the average counts per second [cps] in each energy bin. }
\label{fig:oct10x2023_pn_nu_lc}
\end{figure}
Figure \ref{fig:LIGHTCURVE} shows the $0.3-10$  \kev\ PN light curve and the $4-30$ \kev\ \nustar\ light curve binned to 1000\,s and by orbit (5780\,s), respectively. In the second half of the \nustar\ observation, the $4-30$ \kev\ count rate jumped from $\approx0.2$ to nearly $1$ cps in approximately 30\ks. 

Although \xmm\ was not pointed at the source during the  \nustar\ flare, a smaller brightening event was captured by both telescopes at around 60\ks\ into the \xmm\ observation. Figure \ref{fig:oct10x2023_pn_nu_lc} shows the PN light curve in five energy bands: $0.3-$1 \kev, $1-3$ \kev, $3-4$ \kev, $ 4-6$ \kev, and $6-10$ \kev, all binned to 2000\,s, as well as the \nustar\ low state light curves in the $4-6$ \kev, $6-10$ \kev, and $10-30$ \kev\ band. 

The variability, particularly between $60-85$\ks, is constrained to the bands between $3-30$ \kev. Below 3 \kev, there is no evidence of any variability. The soft light curves here remain constant for the duration of the observation. The  PN $0.3-1$ \kev\ and $1-3$ \kev\ bands have $\chidof=95/91$ ($p=0.37$) and $72/91$ ($p =0.93$), respectively, when compared with their mean. 

When compared to a constant, the  $3-4$ \kev, $4-6$ \kev, and $6-10$ \kev\ light curves resulted in a $\chidof=202/91$ ($p <0.0001$), $344/91$ ($p <0.0001$) and $157/91$ ($p <0.0001$), respectively. Similarly, the  \nustar\ light curves are also inconsistent with constants with $\chidof=86/18$ ($p <0.0001$), $97/18$ ($p <0.0001$) and $82/18$ ($p <0.0001$) for the  $4-6$ \kev,  $6-10$ \kev, and $10-30$ \kev\ bands, respectively. Thus, we reject the null hypothesis that the light curves are constant in all energy bands above 3\,\kev\ with more than 99.99\% confidence. This further indicated that changes to the continuum flux do not propagate to the low energy band on short-time scales.
\begin{figure*}
\centering
\includegraphics[width=\linewidth]{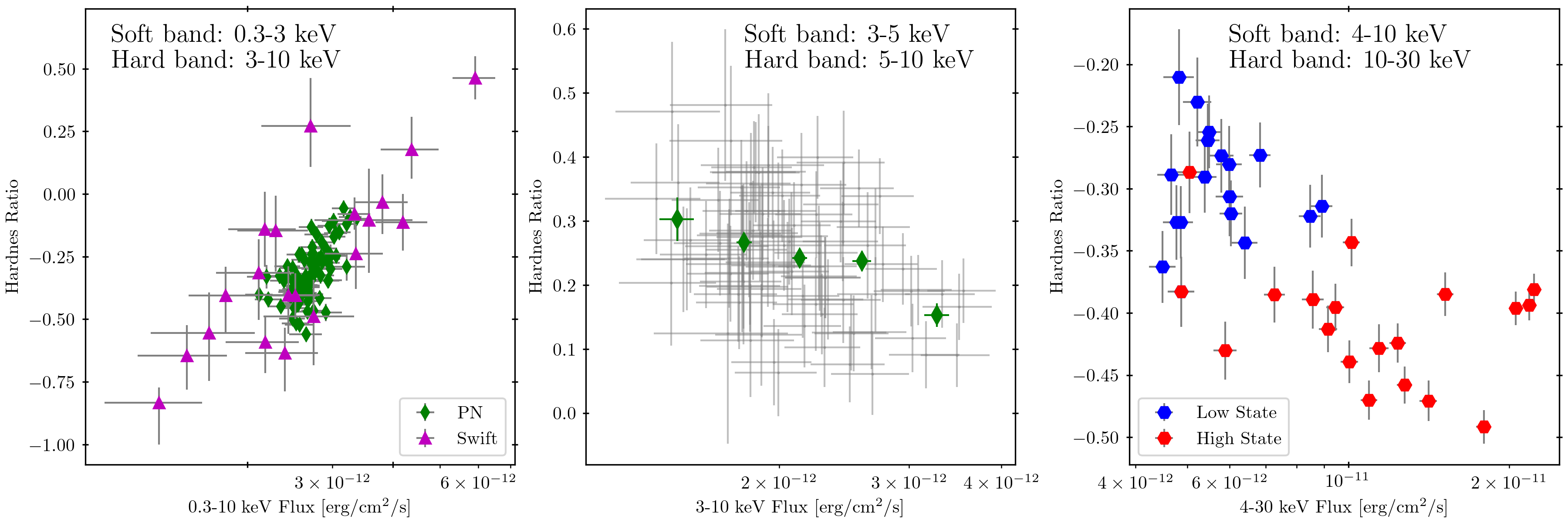}
    \caption{Hardness-Flux plots for the PN \swift\ and \nustar\ data. \textit{Left panel:} \swift\ (magenta triangles) and PN (green diamonds) data Hardness-Flux plot. The hard band is taken to be $3-10$ \kev, and the soft band is $0.3-3$ \kev.  We see a clear, harder-when-brighter trend in the pointed PN observation and the \swift\ monitoring campaign. \textit{Middle panel:} Hardness-Flux plot for PN data, here the hard band is $5-10$ \kev, and the soft band is $3-5$ \kev. The gray data is un-binned, while the green diamonds show the data binned by flux.   \textit{Right panel:} Hardness flux plot for the low state (blue hexagons) and high state (red hexagons) data. The hardness ratio used a hard band of 10-30 \kev\ and a soft band of 4-10 \kev.  Here, we see the expected soft when brighter behaviour. For all \nustar\ data presented, FPMA and FPMB are summed.}
\label{fig:colour-colour}
\end{figure*}  

Hardness-flux plots were constructed for \swift, \xmm\ and \nustar\ data to further demonstrate this disconnect. The left panel of Figure \ref{fig:colour-colour} shows the hardness-flux plot for the PN and \swift\ data with a soft band of $0.3-3$ \kev\ and a hard band of $3-10$ \kev. The \swift\ data have the same time binning as outlined in Table \ref{tab:SwiftObs}, and PN data are again in 1000\,s bins. The average flux was estimated using the average spectra of each data set; this was used to convert cps to flux. The data shows a clear harder when a brighter trend on short (PN) and long (\swift) time scales.  We use Spearman's rank correlation coefficient ($r_s$) to evaluate correlations between the hardness ratio and flux. The $r_s$ and the associated p-value\footnote{ From here the null hypothesis is that there is no correlation between the variables} are calculated using {\sc scipy}\footnote{https://docs.scipy.org/} \citep{Virtanen+2020}. Similar to the Pearson correlation coefficient, positive values of $r_s$ indicate a positive correlation, and negative values of $r_s$ indicate a negative correlation. 

 The $0.3-10$\,\kev\ \swift\ and PN light curves show strong positive correlations, with $r_s=0.77$ and $r_s=0.61$, respectively ($p<0.00007$ for both). This indicates that the source becomes harder as the flux increases.   The flux range for \swift\ is much larger than for PN; we might expect the PN data to continue to follow the \swift\ trend with a more variable observation. This harder-when-brighter trend in the $0.3-10$ \kev\ is not seen in most NLS1 but has been reported in \mrk\ \citep{Buhariwalla+2020}. This points to AGN contribution below 3 \kev\ being absent from detection in \mrk.

Above 3 \kev, a hardness-flux plot was constructed for the PN data with a hard band of $5-10$ \kev\ and a soft band of $3-5$ \kev.  Here, $r_s=-0.35$ ($p<0.0006$) for the un-binned data, which indicates a moderate negative correlation between the flux and the hardness ratio. To further explore this connection, we bin the data into five flux bins.  The Spearman's rank correlation coefficient then becomes $r_s > -0.999$, indicating a strong negative correlation ($p<<0.00005$).

The right panel of Figure \ref{fig:colour-colour} shows a hardness-flux plot for the \nustar\ observation.  Here,  $r_s = -0.67$, ($p<0.00001$) indicating a strong negative correlation. Above 4 \kev, \mrk\ exhibits the expected softer when brighter behaviour.

\subsection{Principal Component Analysis }
\label{sec:PCA}
\begin{figure*}
	\centering
	\includegraphics[width=\linewidth]{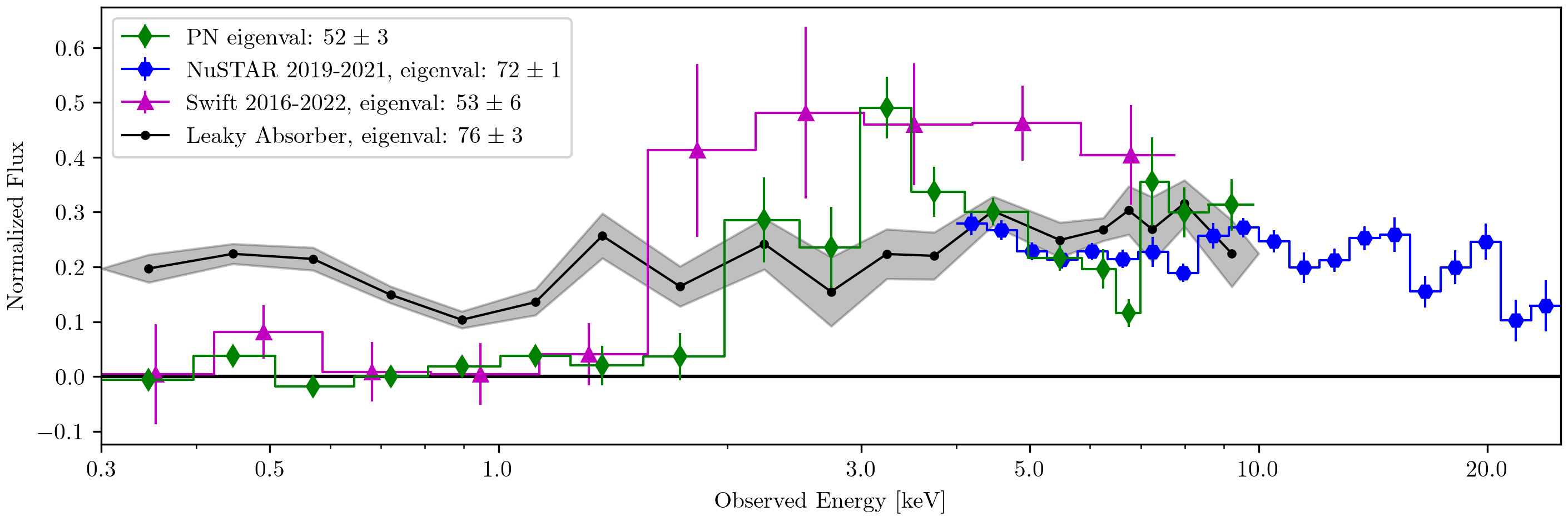}
	\caption{The first principal component for the \swift\ (magenta triangles), PN (green diamonds) and \nustar\ (blue hexagons) as well as the simulated leaky absorber based on \protect\cite{Buhariwalla+2020} in black. The shaded regions on the simulated data show the uncertainties.}
	\label{fig:PCA}
\end{figure*}
Principal Component Analysis (PCA) is a mathematical method to reduce the dimensionality of a multivariate data set. This method deconstructs a set of vectors into a mean vector plus a linear combination of eigenvectors called principal components. After performing PCA, one may reconstruct any input vector by the sum of the mean vector and the principal components scaled to specific values. Each successive principal component will account for less variance than the previous principal component \citep{Ivezic+2019}. The amount of variance captured by each principal component is represented by the eigenvalue; this value can be normalized and presented as a percent variance for each principal component \citep{Parker+2013}.

We utilize time-resolved spectra as our input vectors, allowing us to probe changes in the spectra. This technique has been used for NLS1 galaxies previously (e.g. \citealt{Parker+2013, Parker+2014, Gallo+2015}), and allows us to see the shape of the variable components between spectra.

We perform a principal component analysis on the various data sets presented in the timing analysis. We probe long-term trends with the 20 separate \swift\ spectra; the \nustar\ data were divided into orbit-by-orbit spectra for both the 2019 and 2021 observations.  Finally, the PN data were divided into 20 spectra of approximately 5\ks\ exposure time to probe short-term trends.  

Using these un-binned spectra, we utilized the PCA code distributed by \cite{Parker+2013}\footnote{We port the code from python2 to python3}. We perform the principal component analysis and extract the first principal component and the first eigenvalue. The principal components can be seen in Figure \ref{fig:PCA}. We see that the \nustar\ data suggests a normalization change to explain the variability seen in the observation.  Turning towards the PN and \swift\ principal components. We see the same behaviour as previously reported-- a lack of variability in the soft band and a change in normalization in the hard band. This trend holds on all time scales from the rapid PN to the long-term (\swift) data. The simulated PCA is described in Section \ref{sec:TIMEconc}.

\subsection{Timing Evidence Supporting 100\% Covering Fraction of the AGN Below 3 \kev}
\label{sec:TIMEconc}
\cite{Buhariwalla+2020} presented a continuum model for \mrk\ that required partial covering of the central source. This resulted in emission from the soft band originating from three distinct physical regions: a collisionally ionized component from a star formation region, a photoionized component attributed to distant reflection from the torus, and the partially covered continuum component ``leaking through" the partially covering absorber.  Analysis of the RGS spectra confirmed the presence of the two emission line components, with the detection of lines from collisionally ionized and photoionized emission lines  \citep[see][for details]{Buhariwalla+2023}.  

Utilizing the 2020 best-fit reflection model (see Section \ref{sec:4keV}), using a neutral partial cover with  $\nh=25\times10^{22}\, {\rm \pscm},\, CF=98\%$, we simulate spectral data with variable strengths of the AGN continuum. The continuum strength was varied such that the $0.3-10$ \kev\ count rate ranged between $\sim0.4-0.7$ \cps, consistent with the variation seen in the PN light curves. PCA was performed, and Figure \ref{fig:PCA} shows the first principal component in black. We see a significant amount of variability below $\sim2$ \kev. This is inconsistent with the PN and \swift\ principal components.
\begin{figure}
	\centering
	\includegraphics[width=\linewidth]{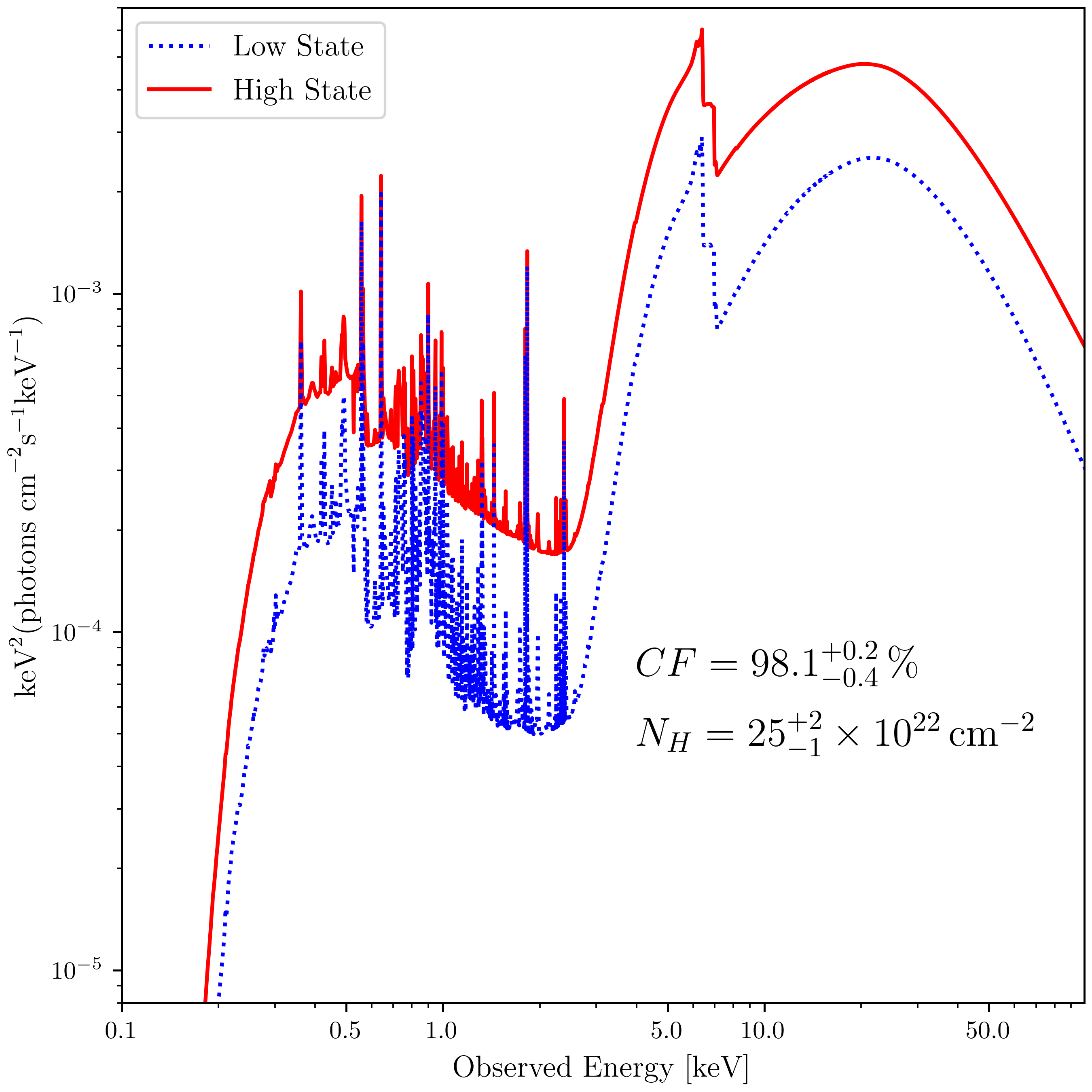}
	\caption{Low state (blue solid line) and High state (red dotted line) best fit partial covering model based on the 2020 best-fit model. The covering fraction of the neutral absorber for both models was $CF=98.2_{-0.4}^{+0.2}$ \%, and the column density was $\nh = 25_{-1}^{+2}\times10^{22}$\pscm\ resulting in $\approx2\%$ of the }continuum leaking through the absorber.
	\label{fig:july19_fig1}
\end{figure}

The origin of the variability detected in the first principal component is primarily due to the AGN continuum leaking through the partial covering absorber. 
This is visualized in Figure \ref{fig:july19_fig1}. Here, the best-fit model from 2020 was applied to the 2021 low-state data. We note that the fit to the 2021 data is acceptable and would be considered one of the best-fit models for the low state.The covering fraction was $CF = 98.2_{-0.4}^{+0.2}$ \%, and the column density was $\nh = 25_{-1}^{+2}\times10^{22}$\pscm. The high-state data was then included, and the AGN continuum components were allowed to scale to match the flux of the high-state spectra. All other parameters (column densities, covering fractions, photon index, etc.) were linked. This is a drastic example of the flux variations seen in the hard band of \mrk\ and is not representative of the flux variation we see in the PN light curve. However, it serves as a visualization of what the hard band of \mrk\ is capable of and how that might affect the soft band of this galaxy.

We further explored the effects of the neutral partial covering material to see if any partial covering absorber could be used to reproduce the first principal component of the PN and \swift\ spectra. We simulated sixteen different absorbers with $\nh=22,\, 25,\, 30,\, 35 \times 10^{22}$ \pscm, and $CF=95,\, 98,\, 99,\, 99.5$\%. For each absorbing scenario, we simulated the spectra such that the  $4-10$ \kev\ count rate ranged between $\sim0.1-0.2$ \cps. This ensured we recovered the variability we saw in the hard band of the PN data. Appendix \ref{D:pca} shows the first principal component for these simulated data. 

None of these absorbers could reproduce the observed first principal component. The results of the PCA simulations indicated that the covering fraction had a stronger influence on the observed variability in the soft band than the column density. However,  all absorbers tested showed soft band variation similar to that shown in Figure \ref{fig:PCA}. Thus, even at covering fractions upwards of 99\%, the continuum leaking through produces significant flux changes in the soft band of \mrk.

The 2020 best-fit reflection model predicts the  $0.3-3$ \kev\ observed flux would be $\num{5.4e-13}$ \ergpscmps\ in the low state, and $\num{1.3e-12}$ in the high state. This would result in an observed change in flux of a factor of $\approx2.4$.  Nowhere in more than 20 years of observations does the flux of the soft band even approach changing that much. The long-term \swift\ light curves show no evidence of \mrk\ exhibiting any soft variability at this level, and the PN light curve showed no evidence of any variability below 3 \kev.

The hardness-flux plot for the \swift\ and PN data showed a clear linear, harder-when-brighter trend (Figure \ref{fig:colour-colour}). All the simulated spectra of \mrk, even with absorbers with parameters at least $5\sigma$\ greater than those required by the low state data ($CF=99.5$\%, $\nh=35\times10^{22}$), fail to reproduce the observed first principal component of the PN spectra.   This suggests that the soft band of \mrk\ is not partially obscured, but  completely absorbed by a $\nh \geq 10^{23}$\cmps\ cloud. This results in no direct AGN emission below 3 \kev.  Instead, the soft band comprises of a  physically large region
that are not intrinsic to the AGN.

This interpretation would explain why the soft band remains constant in the \swift\ 2022 data while the hard band flux dropped significantly. If the continuum was responsible for a significant amount of flux below 3 \kev, then when the hard band dims, so too should the soft. 

\section{Time Resolved Spectral Modelling}
\label{sec:spec}
All spectral data were optimally binned \citep{Kaastra+2016} and background modelled in {\sc xspec} \citep{Arnaud+1996}. We used C-statistics \citep{Cash+1979} to evaluate the fit quality throughout. The data were divided into two epochs, the low-state ($t<100\ks$) and the high-state ($t>100\ks$), as seen in Figure \ref{fig:LIGHTCURVE}. The low-state includes  PN and \nustar\  data, while the high-state includes only  \nustar\  data.  Slowly changing or constant parameters (e.g. inclination, column density, ionization) were linked across both epochs. Furthermore, model components whose emission appeared primarily in the soft band (such as photoionized and collisionally ionized plasmas, etc.) were also linked between epochs. A cross-calibration constant was applied to the FPMA and FPMB detector in \nustar\ when applicable. The constant was free for FPMB and monitored to ensure it remained within acceptable ranges, as prescribed by \nustar\ FAQ\footnote{\url{https://heasarc.gsfc.nasa.gov/docs/nustar/nustar_faq.html}} and the calibration work done by \cite{Madsen+2015A}.   A Galactic column density of $\nh =\num{4.43 e+20}$ (\citealt{Willingale+2013}) was applied to all models and frozen throughout the spectral fitting. Solar abundance in \cite{Wilms+2000} were used throughout the spectral fitting.

For model comparison,  the corrected Akaike information criterion (AICc) was used \citep{Akaike+1974, Sugiura+1978, Liddle+2007}.  AIC  is defined as  $\textrm{AIC} = 2k - 2\textrm{ln}(\mathcal{L})$ where $k$ is the number of fitted parameters and $\mathcal{L}$ is the maximum likelihood estimator (MLE). The exponential of the C-statistic (C-stat) is related to the MLE ($\mathcal{L} \propto e^{-\textrm{C-stat}/2}$),  thus $\textrm{C-stat}=- 2\textrm{ln}(\mathcal{L})$ \citep{Tan+2012}.  With the addition of the small sample size correction, we will use:
\begin{equation}
    \textrm{AICc}= 2k + \textrm{C-stat} + \frac{2k^2+2k}{n-k-1}\;,
    \label{eq:AICc}
\end{equation} 
where  $n$ is the number of data points fitted. The second term goes to zero as $n$ becomes large. Thus, AICc is always valid \citep{Sugiura+1978}. The difference between the AICc from two models, A and B, $\Delta\textrm{AIC}_{A-B} = \textrm{AICc}_{B} - \textrm{AICc}_{A}$, will be used to compare models. We adopt the values of $\Delta\textrm{AIC}\geq6$ to indicate a significantly better fit. This corresponds to a 95 percent confidence that one model is preferred over the other \citep{Tan+2012}. 
\subsection{The Low Energy Spectrum Below 2 keV}
\label{sec:spec2kev}
\begin{figure}
\centering
\includegraphics[width=\linewidth]{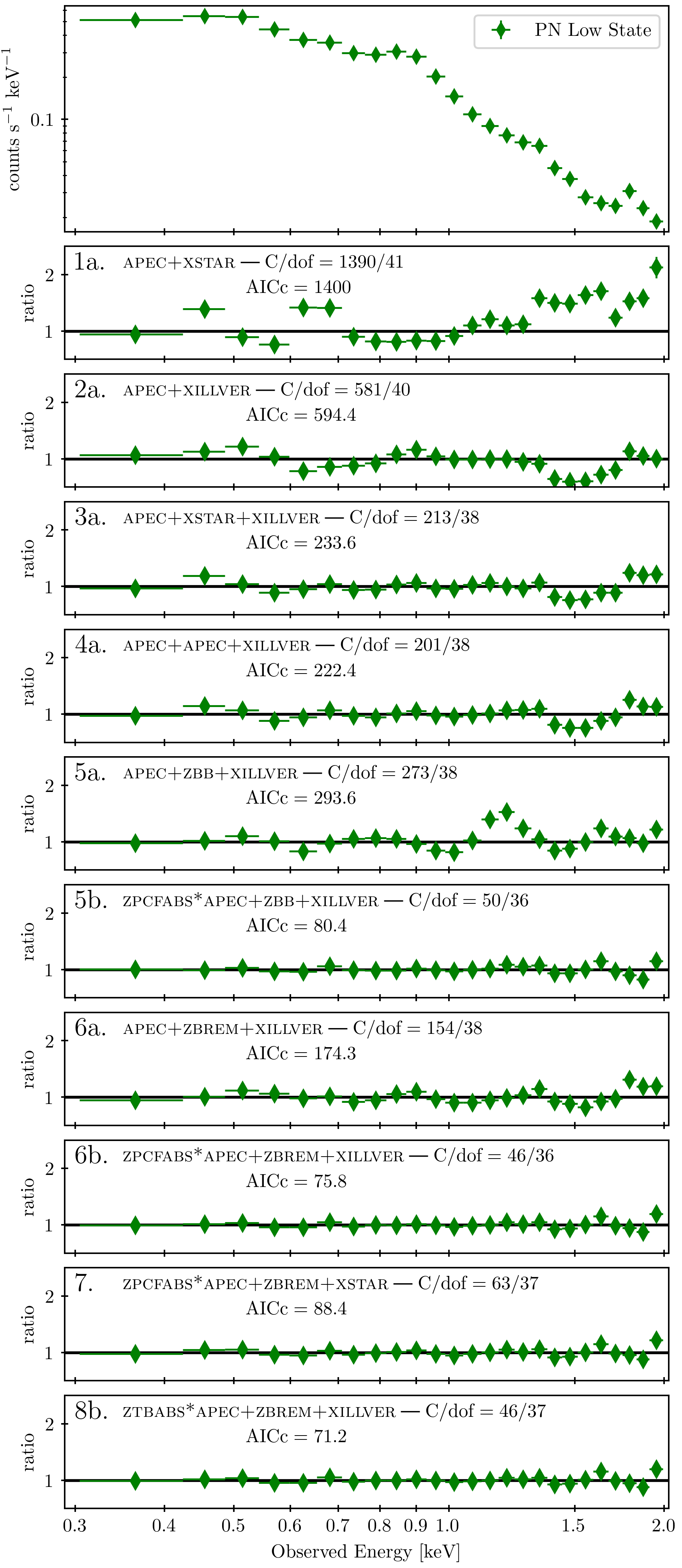}
    \caption{\textit{Top panel:} The $0.3-2$ \kev\ low state PN data. \textit{Lower panels:} Ratio plots from test models for the PN spectra below 2 \kev. The spectral models increase in complexity from top to bottom.  Each model is stated in the panel with its ratio plot along with the \cdof\ and AICc fit statistics. For comparison, all panels share the same energy and ratio axis. The best-fit model can be seen in panel 8b}.
\label{fig:2kev_ratio}
\end{figure}
\begin{table}
\centering
\resizebox{\columnwidth}{!}{%
\begin{tabular}{c l c c }
		\hline
        (1)    & \multicolumn{1}{c}{(2)} 		        & (3)     & (4)   \\
    Model No.  & \multicolumn{1}{c}{Components}         & \cdof\  & AICc  \\
\hline
   	1a	   & {\sc apec + xstar }                    & 1390/41 & 1400  \\ 
   	1b	   & {\sc zpcfabs*apec + xstar }            &  945/39 & 961.8 \\ 
\hline
   	2a	   & {\sc apec + xillver}                   &  581/40 & 594.4 \\ 
   	2b	   & {\sc zpcfabs*apec + xillver}           &  581/38 & 601.5 \\ 
\hline    
        3a	   & {\sc apec + xstar + xillver}           &  213/38 & 233.6 \\ 
   	3b	   & {\sc zpcfabs*apec + xstar + xillver}   &  212/36 & 242.0 \\ 
\hline	
        4a	   & {\sc apec + apec + xillver}            &  201/38 & 222.4 \\ 
   	4b	   & {\sc zpcfabs*apec + apec + xillver}    &  182/36 & 212.0 \\ 
\hline	
        5a	   & {\sc apec + zbb + xillver}             &  273/38 & 293.6 \\ 
        5b	   & {\sc zpcfabs*apec + zbb + xillver}     &   50/36 & 80.4  \\        
\hline
        6a	   & {\sc apec + zbrem + xillver}           &  154/38 & 174.3 \\ 
   	6b	   & {\sc zpcfabs*apec + zbrem + xillver}   &   46/36 & 75.8  \\         
  \hline	
        7      & {\sc zpcfabs*apec + zbrem + xstar}     &  63/37  & 88.4  \\   
  \hline	
        8a     & {\sc zpcfabs*(apec + zbrem) + xillver} &  44/36  & 74.4  \\
        8b     & {\sc ztbabs*apec + zbrem + xillver}    &  46/37  & 71.2  \\
  \hline	
\end{tabular}}
\caption{Models tested with the PN data $<2$ \kev\ that do not contain AGN continuum. Column (1) states the model name, Column (2) shows the components as they would appear in {\sc xspec}, Columns (3) and (4) show the \cdof\ and AICc value for each fit, respectively. The ratios of a selection of these fits can be seen in Figure \ref{fig:2kev_ratio}. }
\label{tab:2kev_AICc}
\end{table}
We begin with the PN spectra below 2 \kev.  As the timing analysis demonstrates, the lack of variability in \mrk\ requires completely removing the varying AGN continuum from the soft band. Thus, we fit the soft band below 2 \kev\ without any AGN continuum. We list each model tested and their fit statistics in Table \ref{tab:2kev_AICc}. Figure \ref{fig:2kev_ratio} showcases the data below 2 \kev\ and a selection of the ratio plots from the models tested. 

We begin with the two-component model from the RGS analysis of  \cite{Buhariwalla+2023}, which utilized {\sc apec} \citep{Smith+2001} for the collisionally ionized component and a generated  {\sc xstar} grid \citep{Kallman+2001} for the photoionized component. The grid was based on those used previously for this source (see \citealt{Buhariwalla+2023}) with density fixed at $10^{10}\,\textrm{cm}^{-3}$, and column density logarithmically sampled at ten points between $10^{20}\,\textrm{cm}^{-2}$ and $10^{24}\, \cm^{-2}$. 

The log ionization was sampled linearly at 10 points between log$\,\xi=1$ and 4 [$\textrm{erg\, cm\,s$^{-1}$}$]. The final grid was calculated at 100 steps, with the energy spanning from 0.1 to 50 \kev\  \citep{Buhariwalla+2023}. For the {\sc apec} component, plasma temperature and normalization were free to vary, while abundance was fixed to solar values. 

This initial model produced a fit statistic of $\cdof=1390/41$. The ratio of this fit can be seen in panel 1a of Figure \ref{fig:2kev_ratio}. This two-component model was expected to fit the PN data poorly as it contains only the emission components visible from the RGS spectra. It is included here to compare with other photoionized components. 

Next, we test a distant reflection component in {\sc xillver} \citep{Garcia+2010, Garcia+2013} for the photoionized component. This model takes an input power law continuum and returns a reflected spectrum based on the viewing angle and physical properties of the material from which the photons are reflected. The  parameters include a photon index ($\Gamma$), high energy cutoff ($E_{cut}$), iron abundance (A$_{\textrm{Fe}}$), ionization (log$\xi$), inclination ($i$) and normalization. Typically, the photon index is linked to the photon index of the primary continuum (power law).  There is no evidence of a power law below 2 \kev; thus, $\Gamma$ was fixed to $\Gamma=2.5$, a typical value of the photon index found in \mrk\ \citep{Buhariwalla+2020, Jiang+2021}. The iron abundance was fixed to solar abundance. The ionization, inclination and normalization were free to vary when {\sc xillver} was implemented. This produced a fit of $\cdof=581/40$. The ratio of this fit can be seen in panel 2a  of Figure \ref{fig:2kev_ratio}.  The ionization parameter was similar to the value obtained through the RGS fitting. This produced a significantly better fit than the {\sc xstar} model, with the $\Delta\textrm{AIC}=806$ indicating that {\sc xillver} is strongly preferred in the two-component scenario. This behaviour is opposite to what we see in the RGS data. We believe that this difference is due to the densities of the photoionized plasma modelled by {\sc xstar} ($n=10^{10}$ ${\rm cm^{-3}}$) and {\sc xillver} ($n=10^{15}$ ${\rm cm^{-3}}$).  This is explored further in Section \ref{sec:PIE}.

Next, we test three component models. The first two components will be {\sc apec + xillver}, and the third will be either a second ionized emission line component ({\sc xstar}, or {\sc apec}) or a smooth continuum-like component such as a black-body ({\sc zbb}) or bremsstrahlung ({\sc zbrem}, \citealt{Karzas+1961, Kellogg+1975}). The {\sc xstar} component will have the same parameters free as above while the {\sc apec}, {\sc zbb} and {\sc zbrem} will have their plasma temperature (kT) and normalization free to vary. The ratios for these three models can be seen in Figure \ref{fig:2kev_ratio} (panels 3a, 4a, 5a, and 6a for  {\sc xstar}, {\sc apec}, {\sc zbb} and {\sc zbrem} respectively). Each model produced a superior AICc than the two-component model, with the data favouring model 6a ({\sc apec + zbrem + xillver}). The $\Delta\textrm{AIC}$ compared to 6a  are as follows: $\Delta\textrm{AIC}_{6-3}=59.3$  compared to model 3a ({\sc apec + xstar + xillver}); 
$\Delta\textrm{AIC}_{6-4}=48.1$  compared to model 4a ({\sc apec + apec + xillver}); and  
$\Delta\textrm{AIC}_{6-5}=119.3$ compared to model 5a ({\sc apec + zbb + xillver}). The bremsstrahlung component appears to be the most favoured third component. 

Models 1 to 6 have corresponding second part (i.e. part b), in which the {\sc apec} component is absorbed with a neutral partial cover ({\sc zpcfabs}). Corresponding models 1, 4, 5 and 6 improve by including the partial cover. The  $\Delta\textrm{AIC}_{a-b}$= 438.2, 10.4, 213.2 and 98.5 for models 1, 4, 5, and 6, respectively. Models 2 and 3 are worse with the inclusion of the neutral partial cover and are only included in Table \ref{tab:2kev_AICc} for completeness.  

For the third component of the soft band model, we find that a smooth continuum component is favoured over a third emission line component. Models 3a and 4b are the best fit three emission line models. However, the  $\Delta\textrm{AIC}>100$, favouring models 5b and 6b, which contain a smooth continuum-like component  ({\sc zbb} and {\sc zbrem}, respectively).

The best model for the PN data below 2 \kev\ requires a smooth continuum-like spectrum in addition to the plasma emission lines.  However, the lack of variability $<2$ \kev\ indicates the third component can not be the AGN continuum.  Here, the shape of the smooth continuum-like component is either blackbody or bremsstrahlung with a preference for the latter ($\Delta\textrm{AIC}_{5-6}= 4.6$). Table \ref{tab:4b_5b} reports the best-fit values for several parameters here,  we note that the error for these parameters was determined using the default {\sc xspec} error command.  The temperature of the bremsstrahlung emission is $\textrm{kT}=0.25\pm 0.03$ \kev, while the temperature of the blackbody was $\textrm{kT}=0.098\pm 0.005$ \kev. The parameters for the neutral partial cover and {\sc apec} components are comparable in all models. Even with the addition of the partial cover on the {\sc apec} component, the temperature of the collisionally ionized plasma is comparable with that measured from the RGS spectra \citep{Buhariwalla+2023}. The temperature of the {\sc zbrem} component is very low, and we would not expect a physical plasma like this to exist. At low plasma temperatures ($<100$ eV), we would expect the two-photon continuum and recombination continuum to dominate over the bremsstrahlung continuum \citep{Bohringer+2010}. Thus, we are using the {\sc zbrem} and {\sc zbb} components to characterize the shape of the spectra and their relative strength.

\begin{table}
\centering
\resizebox{\columnwidth}{!}{%
\begin{tabular}{c c c c c }
		\hline
    (1)           & (2)                       & (3)                     & (4)                   & (5)\\
    Components    & Parameter                 & 5b                      & 6b                    &   8b\\
\hline	
    {\sc zpcfabs} & $\nh$ [$10^{22}$/cm$^{2}$]&$0.74^{+0.14}_{-0.05}$   & $0.77^{-0.05}_{-0.24}$       \\  
     & $CF$ [\%]                   & $>95$                   &$>92$                          \\ 
    {\sc ztbabs}  & $\nh$ [$10^{22}$/cm$^{2}$]&                         &                       & $0.73^{+0.06}_{-0.05}$ \\
    {\sc apec }   & kT [ \kev]                & $0.79\pm0.03$           & $0.80^{+0.04}_{-0.03}$        & $0.80\pm0.03$\\ 
    {\sc  zbb}    & kT [ \kev]                &$0.098\pm 0.005$&\\ 
    {\sc  zbrem}  & kT [ \kev]                &                         &$0.23\pm0.02$ &$0.23^{+0.02}_{-0.03}$             \\          
  \hline	
\end{tabular}}
\caption{Best-fit parameters for  the {\sc zpcfabs} (or {\sc ztbabs}) , {\sc apec}, and  {\sc zbrem} (or {\sc zbb}) components in  models 5b, 6b and 8b Column (1) states the model component, Column (2) states the parameter, and Columns (3), (4) and (5) show the best-fit value for models 5b, 6b, and 8b respectively.}
\label{tab:4b_5b}
\end{table}

We include model 7, {\sc zpcfabs*apec + zbrem + xstar} for completeness. This is the model that we would expect to work best on the RGS data, as the RGS data strongly prefers {\sc xstar} over {\sc xillver}. We can see in Figure \ref{fig:2kev_ratio} that the data are well fit except around 0.5 \kev, which is the location of the \ovii\ triplet. This will be discussed in Section \ref{sec:PIE}.

Finally, we tested the effect of absorption on all three components in model 6. These have been reported in Table \ref{tab:2kev_AICc} as models 8a and 8b. First, we find that the photoionized emission cannot be absorbed. In all absorption tests, the data prefer the photoionized component to be unabsorbed. The fit was significantly worse when all three emission components were fit with the same partial covering component. When all three components have separated absorption, the $\nh$ and covering a fraction of the absorber applied to the photoionized component becomes very small. Next, we find a small fit improvement by absorbing the {\sc zbrem} and {\sc apec} components with the same partial covering component, $\Delta\textrm{AIC}_{6b-8a}=1.4$. This is not at the threshold to be considered statistically significant; thus, we considered these models equivalent. Finally, we test the absorber {\sc ztbabs}, to represent a completely absorbed scenario. We find that under these conditions, the {\sc apec} component can be absorbed, but the {\sc zbrem} component cannot. The C-stat between 6b and 8b is the same; however, {\sc ztbabs} requires one fewer free parameter than {\sc zpcfabs}; thus, the AICc value is smaller. We do not show model 8a  in Figure \ref{fig:2kev_ratio}, as the ratio plot produced is virtually identical to models 6b and 8b.

Model 8b produces the smallest AICc value and will be the model we use moving forward. However, we cannot exclude models  6b and 8a at the 95 percent confidence level. We will revisit the scenarios that produced all three models when we consider the physical interpretation of these results in Section \ref{sec:ISM}.

To explore the validity of these results, we first tested to see if removing all \xmm\ data below 0.5 \kev\ would eliminate the need for a smooth continuum component. This was done to check if possible calibration uncertainties led to the continuum requirements. The smooth continuum component is most present at the softest energies; thus, there was concern that its presence was a statistical artifact. We find that the data still favour the inclusion of the smooth continuum with $\Delta\textrm{AIC}=35.5$, even when all data below 0.5 \kev\ is ignored. Next, we tested if the component was also necessary with the MOS instruments. We find that the inclusion of a smooth continuum component is necessary ($\Delta\textrm{AIC}_{6b-2b}>450$) for the combined MOS spectra. 

Next, we explored how the best-fit model fits the RGS data. We applied model 7 to the data outlined in \citealt{Buhariwalla+2023}. We find that the smooth continuum component is not necessary in the RGS spectra. However, we believe this is due to the dim nature of the source. This will be discussed further in Section \ref{featurlessCont}. Without the need for a smooth continuum, we turn to models 1a and  1b. When they are applied to the RGS data, the $\Delta\textrm{AIC}_{a-b}=19.9$. Thus, we can state that including partial covering of the {\sc apec} component is statistically significant when fitting data from all \xmm\ detectors. The discrepancy between the PN data favouring {\sc xillver} and RGS data favouring {\sc xstar} will be discussed in Section \ref{sec:PIE}.

\subsection{The High Energy Spectrum Above 4 keV}
\label{sec:4keV}
\begin{figure}
	\centering
	\includegraphics[width=\linewidth]{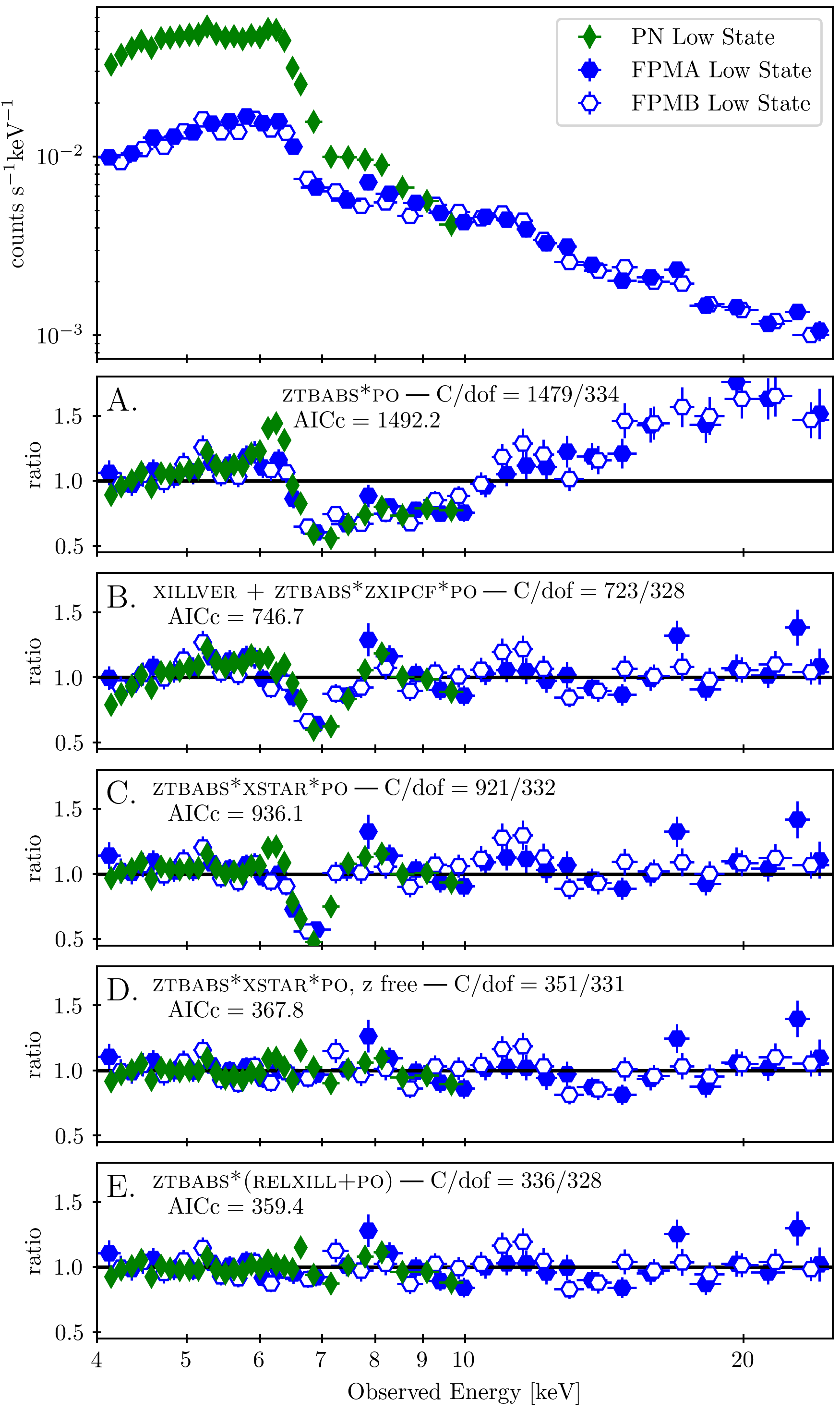}
	\caption{\textit{Top panel}: The PN data between 4-10 \kev, and the \nustar\ low state data between $4-25$ \kev. The PN data are shown with green diamonds while the \nustar\ data are shown in blue hexagons (FPMA is shown with filled hexagons while FPMB is shown with empty hexagons). Panels A, B, C, D, and E show the five models tested.  They include four absorption models and one blurred reflection model. }
	\label{fig:aug23_RA}
\end{figure}

Now, we turn our attention to the spectra above 4 \kev.  Here, the AGN component of \mrk\ can shine through the heavy obscuration that occludes the soft emission. This is also where the \nustar\ flare becomes prevalent.  All models employ a neutral absorber ({\sc ztbabs}) with a column density of at least $\nh=10^{23}$cm$^{-2}$ to ensure that the soft band is completely absorbed. We test several models to account for the AGN emission. Figure \ref{fig:aug23_RA} shows the data and ratios of a sample of these fits.  All parameters are linked between the PN and \nustar\ data for all fits. Cross-calibration constants were free to vary between the FPMA and FPMB instruments.  Despite the known calibration issues between  \xmm\ and \nustar\ above 6 \kev\ (see TN230 for details), we keep the response the same and link $\Gamma$ between \nustar\ and PN data. This is because inadequacies in the models, especially in the early testing phase, far outweigh cross-calibration issues with the data. 

The ratio of an absorbed power law fit can be seen in panel A of Figure \ref{fig:aug23_RA}. This does not describe the broad feature between 6 and 7 \kev. There may also be an emission line feature around 6.4 \kev.  Furthermore, there is an excess of emission above 10 \kev\ that appears to peak at $\sim20\,\kev$, possibly suggesting the presence of a Compton hump. 

Next, we explored the preferred continuum scenario from \cite{Buhariwalla+2020}, Ionized Partial Covering (IPC), which can be seen in panel B of Figure \ref{fig:aug23_RA}.  In this scenario, the curvature seen in the spectra is a result of absorbing material partially obscuring the central engine, blocking some fraction of the emission while letting the rest pass through the material (see \citealt{Holt+1980, Tanaka+2004, Buhariwalla+2020}). 

The original model used for \mrk\ was {\sc xillver + zxipcf*zpcfabs*po}. We have shown that the absorption of the soft band is complete; thus, the neutral partial cover ({\sc zpcfabs}) is replaced with neutral absorption ({\sc ztbabs}). The  {\sc xillver}  component is included to model the neutral \feka\ line at 6.4 \kev. This was an improvement over the absorbed power law. However, the curvature of the spectrum was not easily reproduced with absorption. The model over-predicts the emission at $\sim6.7\, \kev$.  
We then attempted to fit the data with as many absorbing (ionized and neutral) components as necessary to test if a complex absorbing scenario was possible. We could not obtain a fit with traditional absorption models that were statistically better than the IPC scenario in panel B. 

Panel C of Figure \ref{fig:aug23_RA} shows the $4-25$ \kev\ using a photoionized absorber {\sc xstar} grid. This grid was constructed with properties described in Section \ref{sec:spec2kev} but with $A_{\textrm{Fe}}=10$.  With this fit, there is a deep residual at 6.7 \kev, where the model drastically over-predicts the data. This can be solved by allowing the redshift of the {\sc xstar} component free to vary. This produces the best-fit absorption model for these data. This is the only absorption model that could reproduce the shape of the spectra between 6 and 7 \kev.  However, this model requires the absorbing material to be in-flowing at $\sim0.07$c.   

Similarly, if we replace the neutral reflector with an absorbed {\sc pexrav} component \citep{Magdziarz+1995} and also allow it to inflow, it produces a similar quality of fit as panel D (${\rm AICc}=372.8$, ${\rm C-stat/dof}=349/328$). More interesting is that the inflow velocity is $\sim0.07$c, the same as in the absorbing scenario. This is attributed to the prominent edge feature produced at $\sim7.1$ \kev\ by both the absorption and reflection spectra. This edge feature is redshifted to align with the steep drop in the hard spectrum of \mrk\ observed near 6.7 \kev.  

Material inflowing at these velocities is a highly unlikely scenario, as \mrk\ has a high Eddington ratio ($\lambda=0.2-2$, \citealt{Pan+2021, Jiang+2021, Gravity+2023}). Thus, we would expect any material in our line of sight to be outflowing rather than inflowing, \citep[but see e.g.][]{Pounds+2018}. This scenario represents the only way these data can be fit absorption. 

The final model tested for the hard band is a blurred ionized reflector (see \citealt{Ballantyne+2001}; \citealt{Miniutti+2004}; \citealt{RossFabian+2005}). This is a reflection spectrum that is subject to the relativistic blurring effects near the SMBH. The most often implemented model is {\sc relxill} \citep{Garcia+2014, Dauser+2013}. This model contains all the same parameters as {\sc xillver} (photon index, $\Gamma$; iron abundance, $A_{\textrm{Fe}}$; disc ionization, log$\xi$, and inclination) plus and blurring parameters like the inner disc radius ($r_{in}$) and outer disc radius ($r_{out}$), black hole spin ($a = cJ/GM^2$, where $M$ is the black hole mass and $J$ is the angular momentum) and the emissivity profile. The emissivity profile dictates the disc illumination as a function of distance ($r$) and is dependent on the geometry of the inner disc and the corona.  It is described by a broken power law ($\propto  r^{-q}$) with index 1 ($q= q_{in}$) used prior to the break radius ($r_{br}$) and index 2 ($q=q_{out}$) after. We define the reflection fraction as the observed ratio of reflected to continuum flux between $0.1-100$ \kev. This scenario is often used to describe the spectral and timing properties of NLS1s (see \citealt{Fabian+2009B}; \citealt{Chiang+2015}; \citealt{Gallo+2013}, \citealt{Gallo+2015}; \citealt{Jiang+2019}; \citealt{Waddell+2019}). In \cite{Buhariwalla+2020}, the data were insufficient to constrain a realistic blurred reflection model. Specifically, the emissivity profile could not be constrained, yielding results that were difficult to interpret. The inner and outer emissivity index were fixed to a constant value of 3.

We have sufficient data quality to allow the emissivity profile to vary. However, initial tests find the break radius and $q_{out}$ are degenerate. This is somewhat unsurprising as the emissivity profile is very difficult to constrain. The value of $q_{out}$ we recover was consistent with the Newtonian value of $q_{out}=2$, indicating the primary X-ray source is at a large distance. The break radius is sufficiently high to be consistent with this scenario, so we fix  $q_{out}=2$.  We test fixing $q_{out}=3$ and find the AICc value increases by $\Delta\textrm{AIC}=7.3$, indicating these values produce a similar result, but $q_{out}=2$ is preferred with more than 95 per cent confidence. We note that when $q_{out}=3$, the $q_{in}$ value produced is consistent with what we report here, but the break radius becomes much smaller ($\sim10$\rg). 

Initial fits always pegged the high energy cut-off at maximum, so this parameter was fixed to 300 \kev. The inner radius is fixed at the innermost stable circular orbit, and the outer radius is fixed at 400\rg. The spin of the black hole is fixed to $a=0.998$. The reflected photon index is linked to the intrinsic power law photon index, which is allowed to vary. The inclination, ionization parameter and iron abundance are also all free to vary. 

The reflection scinario sufficiently models the spectral shape of the hard band, producing an $\textrm{AICc}= 359.4$. This is a better fit than the inflowing absorption model ($\Delta\textrm{AIC}_{D-E}=8.4$), and it invokes a much more plausible scenario than ionized material in-flowing at 0.07c.

\subsection{Fitting the broadband, Pre-Flare Low State Spectra}
\label{sec:lowStateSpectra}
\begin{figure}
	\centering
	\includegraphics[width=\linewidth]{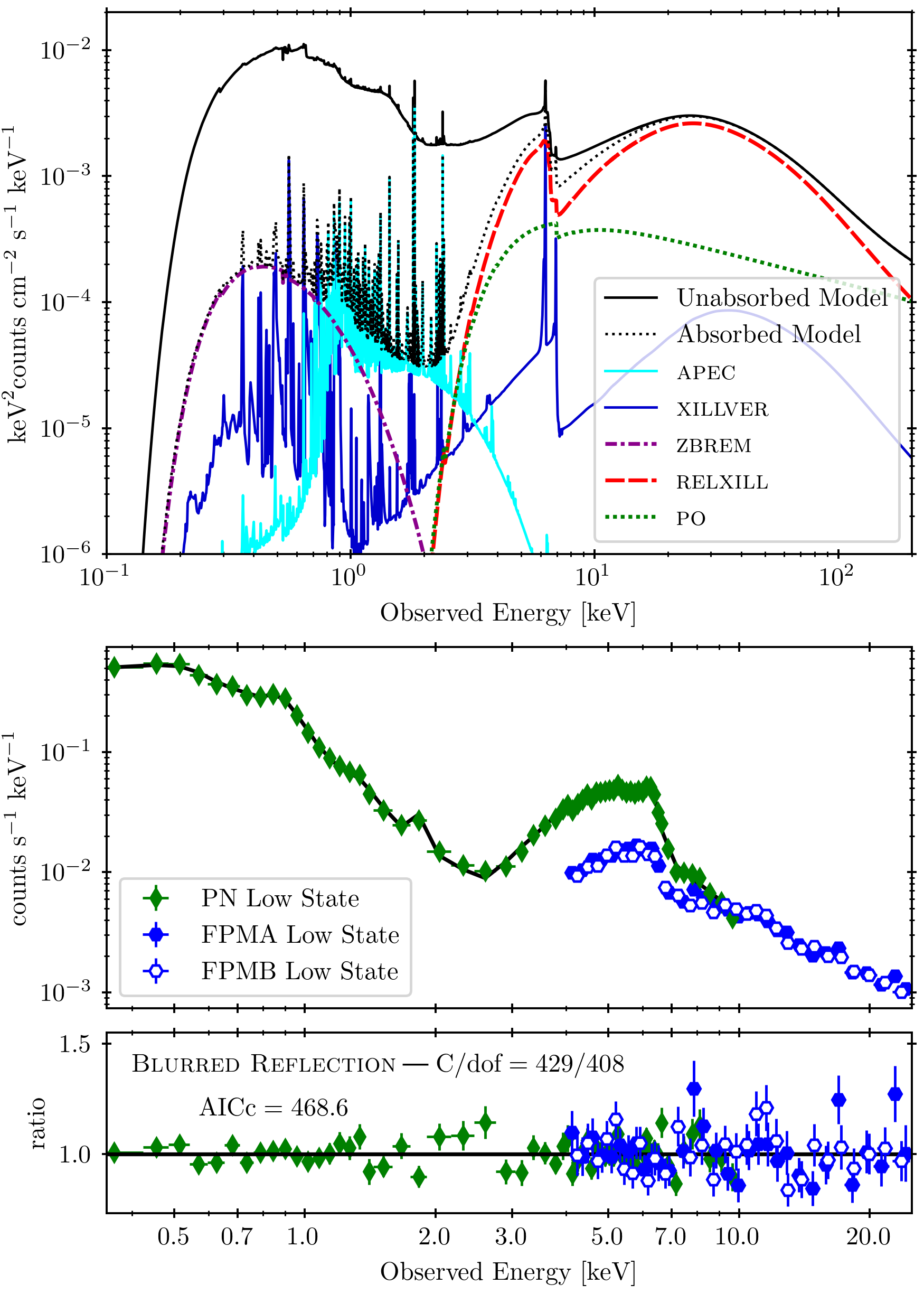}
	\caption{\textit{Top panel:} The $0.1-200$ \kev\ low state model. Black solid and dotted lines show the unabsorbed and absorbed models, respectively. The individual emission components of the models are also shown. The bremsstrahlung ({\sc zbrem}) component is shown with the dashed purple line, the photoionized component ({\sc xillver}) is shown in dark blue, while the collisionally ionized component ({\sc apec}) in light blue. The  AGN continuum is comprised of the primary emitter ({\sc po}) in the dotted green line, and the blurred reflection component ({\sc relxill}) is shown with the dashed red line. \textit{Middle panel:} The low state data. The $0.3-10$ \kev\ PN data is shown in green diamonds, the $4-25$ \kev\ \nustar\ spectrum is shown in blue hexagons  FPMA and FPMB are shown with filled and unfilled hexagons, respectively.  \textit{Bottom panel:} The ratio of the blurred reflection fit, same labels as the middle panel. The data has been binned for clarity.}
	\label{fig:pl_eem_ld_ra_low}
\end{figure}

Combining the best-fit soft band model with the best-fit hard band model allows us to explore how they might work together. The broadband model reads as {\sc ztbabs*apec + zbrem +xillver+ ztbabs*(po + relxill)} and was fit across the $0.3-10$ \kev\ PN spectrum together with the $4-25$ \kev\ \nustar\ spectrum. The inclination was linked between {\sc xillver} and {\sc relxill}, and the reflection photon index for all components was tied to the power law photon index. The fit statistic for this model was $\cdof=429/408$, ${\rm AICc}=468.6$. Figure \ref{fig:pl_eem_ld_ra_low} shows the data, models and ratios. 

We also tested different flavours of {\sc relxill} models, including {\sc relxilllp} for reflection with a lamppost geometry and {\sc relxillCp}, for a variable density accretion disc and Comptonized continuum \citep[see][for details]{Dauser+2022}. We find a worse fit for both, with a $\Delta\textrm{AIC}=11.7$ and  $\Delta\textrm{AIC}=4.9$ for {\sc relxilllp} and {\sc relxillCp}, respectively. Thus, we continue with the standard flavour of relxill.  

We tested whether un-linking the photon index between PN and \nustar\  affected the fit statistic. The  $\Delta\textrm{AIC}=2.1$, favouring the linked photon index. If we link the iron abundance between {\sc xillver}, and {\sc relxill}, the fit is worse by $\Delta\textrm{AIC}=6.5$. The reflection fraction in the low state was calculated as the ratio between the blurred reflected flux and the power law flux between $0.1-100$ \kev. It was found to be $R = 2.4_{-0.5}^{+1.3}$. We note that \cite{Buhariwalla+2020} struggled to find the reflection fraction with the available data because of the lack of power law contribution below 10 \kev. In the absence of data below 10 \kev, the power law component is negligible in the models, and only an upper limit can be estimated for the reflection fraction.

We also tested whether having the spin, inner disc radius, and cutoff temperature of the corona free affected the fit; they did not. The spin is constrained to $a>0.6$, and leaving it free to vary did not improve the fit. Similarly, the inner radius was constrained to be $r_{in}<4r_g$. Having the inner radius free did not impact the fit statistics.  The cutoff temperature remained unconstrained. We thus left the spin, inner radius and cutoff temperature fixed to the values listed in Table \ref{tab:relxill}.

We note there are intrinsic difficulties constraining reflection parameters without soft band data \citep[see][for details]{Bonson+Gallo+2016}. Due to the nature of \mrk, the intrinsic AGN emission in the soft band data is completely absorbed below $3\,\kev$. Thus, we interpret the parameters carefully. As per \cite{Bonson+Gallo+2016}, and similar works \citep{Choudhury+2017, Kammoun+2018}, parameters such as photon index and inclination are relatively easy to measure. However, spin, $q_{in}$ and log$\xi$ are more difficult to recover accurately. Instead of focusing on the value of individual parameters, we will broadly interpret the physical scenarios.

Overall, the soft and hard band models come together quite well, considering they were developed independently.  An interesting note about this model is that when the $10^{23}$cm$^{-2}$ absorber is removed from the AGN continuum (solid black like in Figure \ref{fig:pl_eem_ld_ra_low}), much of the interesting soft band behaviour of this source is washed away.  Simulating an 80 \ks\ PN spectrum using this model can adequately fit the data using blurred reflection and a power law component, eliminating the need for extra soft band components. The simulated data, along with the observed data, can be seen in Appendix \ref{B:PN}. 

\subsection{Self-Consistent Spectral Fits including the flaring High-State}
\begin{figure*}
	\centering
	\includegraphics[width=0.95\linewidth]{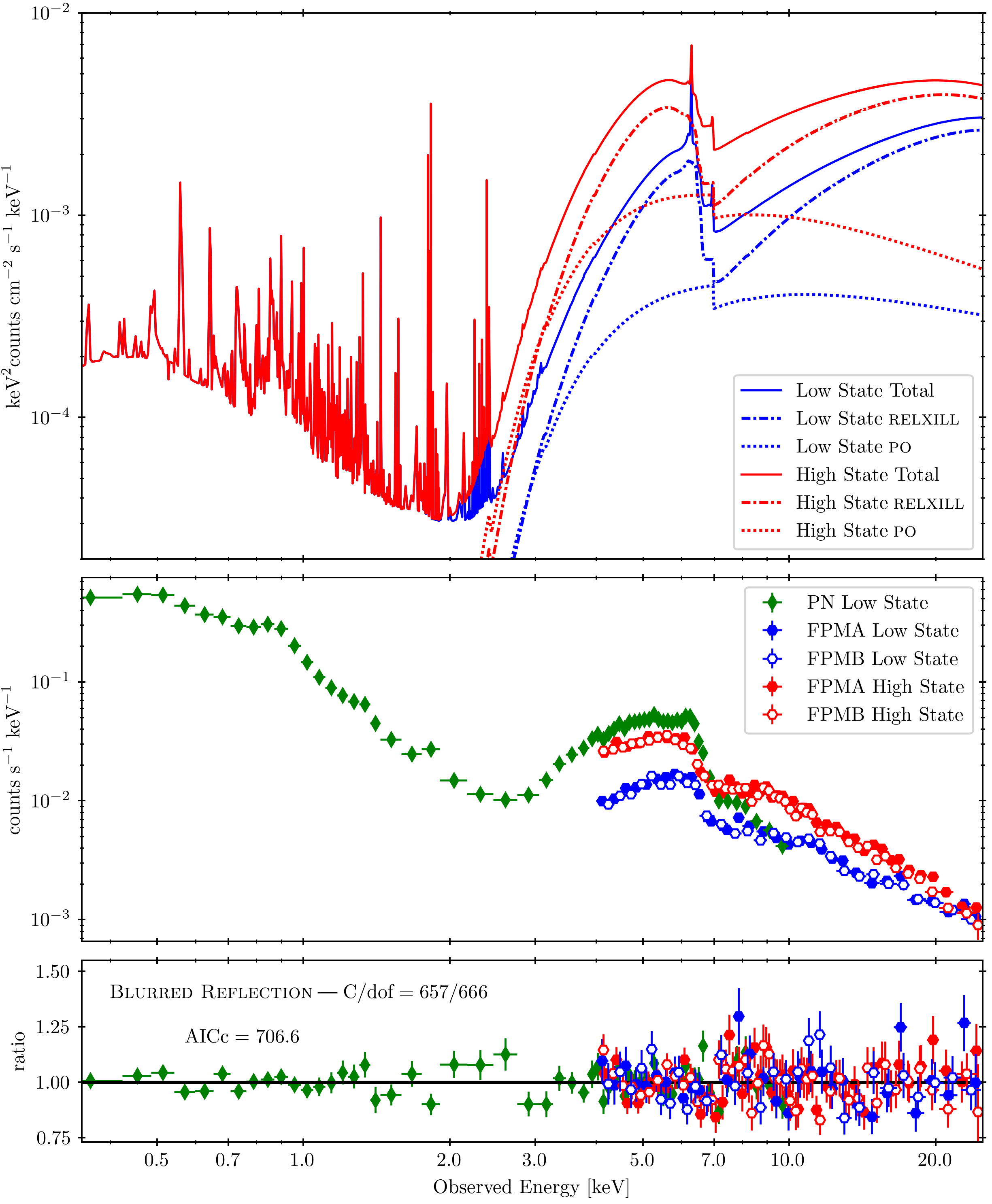}
	\caption{\textit{Top panel:} The $0.3-25$ \kev\ total model for the low and high states are shown in the solid blue and red lines, respectively. The low and high state reflection (dot-dashed line) and power law (dotted line) components are also shown, as these are the only components that show variability between the low and high states. \textit{Middle panel:} The low and high state data. The  $0.3-10$ \kev\ PN data is shown in green diamonds, the $4-25$ \kev\ \nustar\ low state spectrum is shown in blue hexagons, and the $4-25$ \kev\ \nustar\ high state spectrum is shown in red hexagons.  FPMA and FPMB are shown with filled and unfilled hexagons, respectively.  \textit{Bottom panel:} The ratio of these data fit with the blurred reflection model, with the same labels as the middle panel. The data has been binned for clarity. }
	\label{fig:pn_nuL_nuH_spec}
\end{figure*}

Next, we attempt to fit the 2021 high-state spectra with the best-fit model to the low-flux state (Section \ref{sec:lowStateSpectra}). Parameters that we expect to be constant or slowly varying were linked between the low and high states.  These include all soft-band components, the spin, inclination, and ionization of the accretion disc. We allowed parameters that we expect to respond quickly free to vary. This includes the photon index, emissivity profile and normalization of the power law and reflection component. We find that the high state can be well fit by linking $q_{out}$ and the break radius, allowing $q_{in}$, $\Gamma$, and normalization free.  Figure \ref{fig:pn_nuL_nuH_spec} shows the best-fit models, data and ratios of this fit, while Table \ref{tab:relxill} shows the best-fit parameters. The corner plots for the MCMC error calculation can be seen in Appendix \ref{C:mcmc}. 
This model produced a fit of $\cdof= 657/666$, ${\rm AICc}=706.6$. We also note the best-fit absorption model can not describe the 2021 flare with only photon index and normalization changing and has a $\Delta{\rm AIC}=66.5$, and the absorber must be inflowing at $\sim0.11$c for the high state (the low state still required inflow at $0.07$c). 

\begin{table*}
\centering
\begin{tabular}{c c c c c}
\hline
(1)                &      (2)        &        (3)            &     (4)              &  (5)                   \\
Physical Component & Model Component & Model Parameter       & Low State            & High state             \\
\hline
Neutral absorber   & {\sc ztbabs}  & $\nh$\ [$10^{22}$/cm$^2$] &$0.73\pm 0.05$      &                        \\
\hline
Collsionally Ionized& {\sc apec}   & kT [keV]                & $0.80\pm 0.03$       &                        \\
Component          &               & Abundance [solar]       & $1^f$                &                        \\
                   &               & norm	                  & $3.2\pm 0.3(\times10^{-4})$&                  \\
\hline
Bremsstralung      & {\sc zbrem}   & kT [keV]                & $0.23\pm0.02$	    &                        \\
Emission           &               & norm	                 & $1.6\pm 0.2(\times10^{-3})$&                  \\
\hline
Photoionized       & {\sc xillver} & log$\xi$ [\ergpscmps]   & $1.21^{+0.11}_{-0.12}$         &                        \\
Emission           &               & Inclination [$\deg$]    & $17^l$	            &                        \\
                   &               & A$_{\textrm{fe}}$	[solar]& $1^f$	             &                        \\
                   &               & $\Gamma$                & $2.44^l$             &	                      \\
                   &               & norm	                  & $5\pm 1(\times10^{-6})$&                  \\
\hline
Neutral absorber   & {\sc ztbabs}  & $\nh$\ [$10^{22}$/cm$^2$]& $29.2^{+1.3}_{-1.4}$     &                        \\
\hline
Blurred reflector  & {\sc relxill} & $q_{in}$                & $3.53\pm 0.11$       & $4.02_{-0.15}^{+0.16}$ \\
                   &               & $q_{out}$               & $2^f$            	 &                        \\
                   &               & $r_{Br}$ [\rg]          & $46_{-7}^{+11}$	     &                        \\
                   &               & $r_{in}$ [\rg]          & $1.2^{f,*}$	         &                        \\
                   &               & $r_{out}$ [\rg]         & $400^f$	             &                        \\ 
                   &               & spin                    & $0.998^f$            &                        \\
                   &               & Inclination [$\deg$]    & $17_{-6}^{+4}$	     &                        \\
                   &               & log$\xi$ [\ergpscmps]   & $2.6\pm 0.2$         &                        \\
                   &               & A$_{\textrm{Fe}}$	[solar]& $>9$              &                        \\
                   &               & $E_{cut}$               & $300^f$              &                        \\
                   &               & $\Gamma$                & $2.43^l$             &	$2.83^l$              \\
                   &               & norm	      &$4.1_{-0.9}^{+1.0}(\times10^{-4})$&$39_{-10}^{+17}(\times10^{-4})$\\
\hline
Intrinsic          & {\sc powerlaw}& $\Gamma$                &$2.44\pm0.07$         & $2.82_{-0.04}^{+0.09}$ \\
Power Law          &               & norm	                 &$1.5\pm 0.4(\times10^{-3})$&$8.9_{-1.3}^{+1.4}(\times10^{-3})$\\
\hline
Calibration Constatnts& FPMA       &                         & $1.01_{-0.03}^{+0.04}$        &                        \\
                   & FPMB          & 	                      & $1.03\pm0.04$	     & $1.02\pm0.03$          \\
\hline 
\hline
\end{tabular}
\caption{Best fit parameters for the low and high state models. Column (1) lists the physical description of the {\sc xspec} component listed in Column 2. Column (3) outlines the model parameters from each model component.  Columns (4) and (5) show parameter values for the low state and high state, respectively. If a parameter is blank in Column (5), that value is linked between the low and high states. The parameters with superscript $f$ have been fixed to the displayed value, while the parameters with the superscript $l$ have been linked to the same parameter in a different model component within the model. ($*$) This is fixed to the innermost stable circular orbit, which for a maximal spinning black hole is 1.2\rg.}
\label{tab:relxill}
\end{table*}

The low state inner emissivity index was measured to be $q_{in,L}=3.53\pm 0.11$, while the high state was measured to be $q_{in,H}=4.02_{-0.15}^{+0.16}$. We tested to see if this difference was due to the nature of the instruments used for the observation. This was done in two ways. First, we tested whether allowing $q_{in}$ for the low state \nustar\ spectra to be separate from the low state PN spectra produced a significantly better fit. This gave  $\Delta\textrm{AIC}=4.0$, favouring the separate indices. However, the separate $q_{in}$ values were comparable. Next, we linked all the emissivity profile parameters between the low and high states. This produced a significantly worse fit, with a $\Delta\textrm{AIC}=80.3$, indicating that separate inner indices are strongly preferred between the low and high states. Together, this suggests that the change in the emissivity profile is required by the data rather than a result of the different instrumentation.

While it is difficult to state properties of the accretion disc with these data confidently, we can infer some things about the regions that produced the reflection spectra. Based on the emissivity profile, we are seeing limited light bending effects as compared to other sources like MCG-6-30-15 \citep{Brenneman+2006} and 1H\,$0707-495$ \citep{Fabian+2012A}, which have steep inner emissivity profiles (e.g. $q_{in}>6$).  Instead, our value is closer to $q_{in}\sim4$, indicating we are viewing reflection off the accretion disc at some distance from the SMBH. We do not speculate why as there could be a myriad of factors (different spin, inner radius, absorption effects, etc.); we only speculate on why the inner emissivity index may have changed during the flare.     

Overall, our analysis demonstrates the differences between the low-flux and flaring levels can be described by changing the reflection fraction and overall flux of the power law. The low state reflection fraction was measured to be $R_L=2.4_{-0.5}^{+1.3}$, while the high state reflection fraction was measured to be $R_H=1.6_{-0.4}^{+0.6}$.  The low state photon index was measure to be $\Gamma_L=2.43\pm0.06$ and the high was $\Gamma_H=2.82_{-0.04}^{+0.09}$. The physical motivations of this flare will be discussed in the next section.

\section{Spectral Variability Analysis}\label{sec:spec_var}
\subsection{Principal Component Analysis Based on New Best-fit X-ray Model}
In Section \ref{sec:PCA}, we established that the best-fit ionized partial covering model from \cite{Buhariwalla+2020} could not adequately describe the variability of \mrk\ despite providing a good fit to the spectrum. Here, we will perform a principal component analysis based on the best fit spectral model discerned in Section \ref{sec:spec} to determine if it could reasonably describe the rapid and long-term variability. The main difference between the new model and that of \cite{Buhariwalla+2020} is that the covering fraction of the neutral absorber is 100\%, indicating no AGN continuum leaking into the soft band. 

Twenty PN spectra and forty \nustar\ spectra were simulated like in Section \ref{sec:PCA}. The twenty PN spectra had the normalization of the power law and blurred reflection component change such that the resultant spectra had count rates ranging from $\sim0.4-0.7$\cps, consistent with the count rate variation seen in the PN light curves.

For the forty \nustar\ spectra, twenty had power law and reflection normalization changing by $\sim20$\%, representing the low state and the 2019 data. The other twenty had variable power law and reflection normalization, changing by a factor of 5 to represent the flaring data. The $4-30$ \kev\ count rates ranged from $0.15-0.75$ \cps, representing the variability in the \nustar\ light curves. 

\begin{figure*}
\centering
\includegraphics[width=\linewidth]{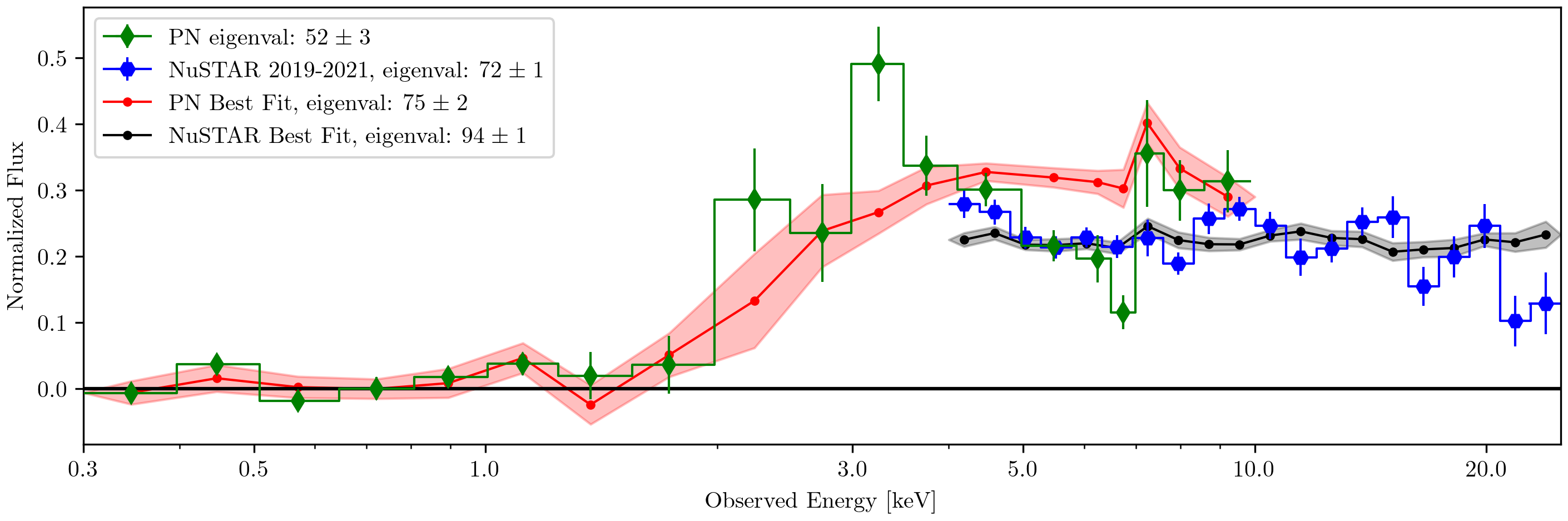}
    \caption{The observed first principal component for the PN and \nustar\ data are shown in green diamonds and blue hexagons, respectively. The simulated first principal component based on the best-fit model for the PN and \nustar\ data is shown with the red-shaded and grey-shaded regions, respectively. We can see that the general trends in variability are better reproduced using this model.}
\label{fig:PCA_fig2}
\end{figure*}

This was somewhat of a naive approach to simulating the spectral variability as we only account for a monotonic change in the AGN continuum. The reflection fraction, photon index, and inner emissivity index remain constant throughout. These are parameters we know to vary during the \nustar\ flare. However, this approach is sufficient for the first-order approximation we are attempting here. Figure \ref{fig:PCA_fig2} shows the first principal component for the PN and \nustar\ data along with the first principal component for the simulated data.

With this model, the simulated PN spectrum matches the observed data well. There is a slight disagreement around 3 \kev\ and around the location of the broad iron line, which could be improved upon by complicating the model (e.g. varying the reflection fraction or ionization parameter). The simulated \nustar\ principal component matches the overall data well, suggesting this simplistic model for the variability is responsible. 

Moreover, the model also reproduces the negligible variability below 3  \kev. The easiest explanation for the consistent nature of the soft band is that it is physically large. This was the interpretation used in \cite{Buhariwalla+2020}, where the CIE originated from regions of star formation within the galaxy. A very long timeline would be needed to explore variations of three ionized components making up the soft band. To attempt to examine the variability of the three soft band components with the current data, we fitted the $0.3-2$ \kev\ PN spectra, the 2001 MOS, 2007 \suzaku\ XIS, and the 2019 \swift\ XRT spectra with the same model described in Section \ref{sec:spec2kev}. All parameters were linked between epochs, and only the normalization of each component was allowed to fluctuate. This was done to explore any possible variation of the strength of each component across the 20 years of data. The $0.2-2$ \kev\ luminosity of each component was measured using {\sc clumin}. The snapshot \swift\  observations were not included in this analysis. 
\begin{figure}
\centering
\includegraphics[width=\linewidth]{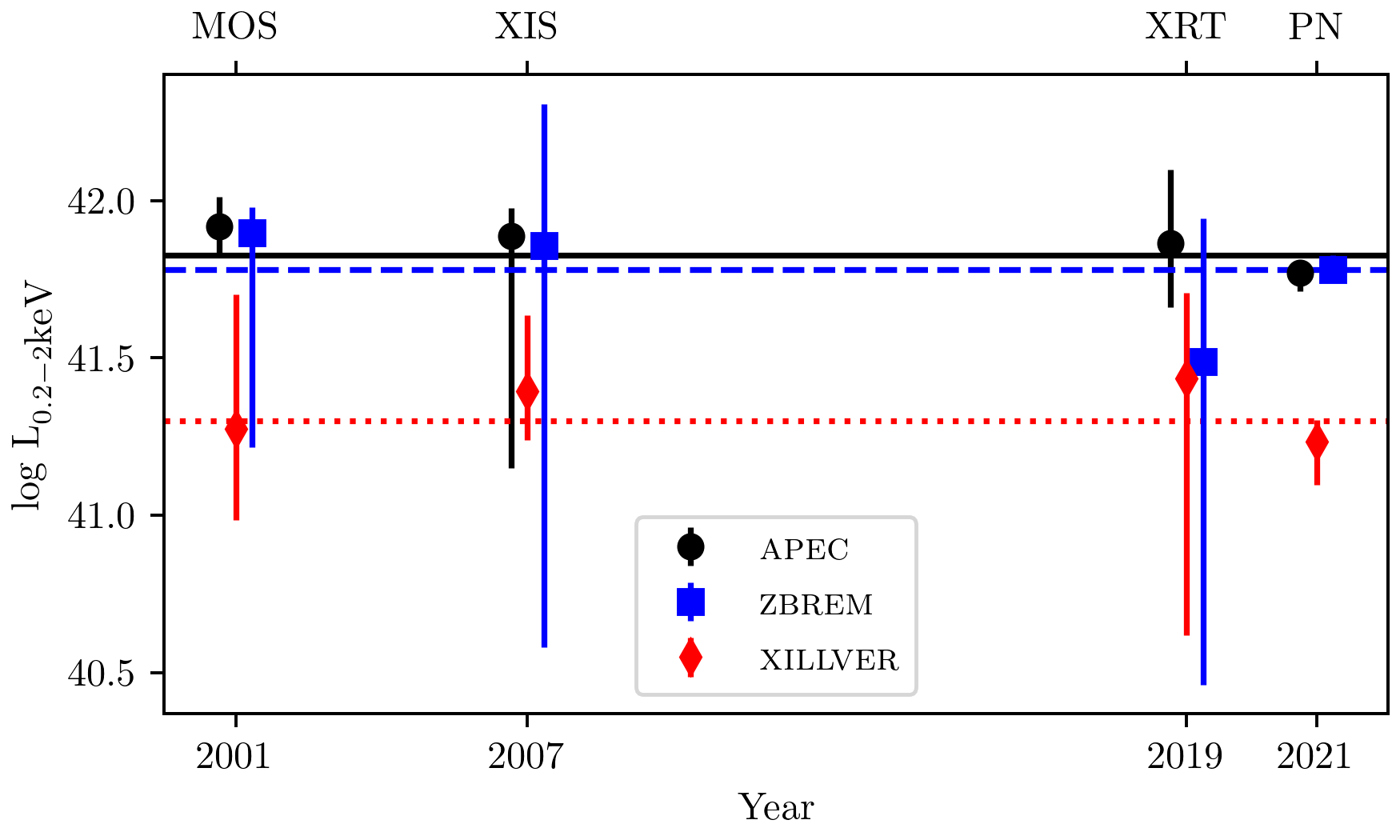}
    \caption{The $0.2-2$ \kev\ luminosity of the {\sc apec} (black circle), {\sc zbrem} (blue square), and {\sc xillver} (red diamond)  component at each epoch that spectral data is available for. The black solid line shows the error weighted mean luminosity for the {\sc apec} component, the blue dashed line for the {\sc zbrem} component and the red dotted line for the {\sc xillver} component. The lower axis outlines the year of the observation, while the upper axis shows the instrument used to observe.  A small horizontal offset in the date is introduced for visual clarity.}
\label{fig:2kev_lumin}
\end{figure}

Figure \ref{fig:2kev_lumin} shows the unabsorbed luminosity for each component of the soft band at each epoch. We 
compare the measurements to the error-weighted mean luminosity over all epochs. We can see that the {\sc apec} and {\sc zbrem} components have roughly the same unabsorbed luminosity. However, when absorbed, the luminosity of the \apec\ component is $\sim0.7$ dex lower, making its observed luminosity the lowest of the three components.  

The {\sc apec} component is the only one well-constrained at every epoch. The reason for this is likely because the three instruments used all have adequate effective area and calibration at approximately 0.9 \kev. This is where the main feature of the {\sc apec} component is evident.  This feature is present above whatever continuum is present in this object.  The {\sc zbrem} is relatively strong in all epochs; however, it is hard to accurately constrain its strength as it peaks at very low energies, and only the high energy tail is observable.

Based on these spectral data, we see no evidence of variability in any of the individual components that comprise the soft band. 
A long timeline and high-quality data would be needed to probe any subtle changes fully.

\subsection{The Rapid Flaring Event Captured by \nustar}
\label{sec:flare}
\begin{figure}
\includegraphics[width=\linewidth]{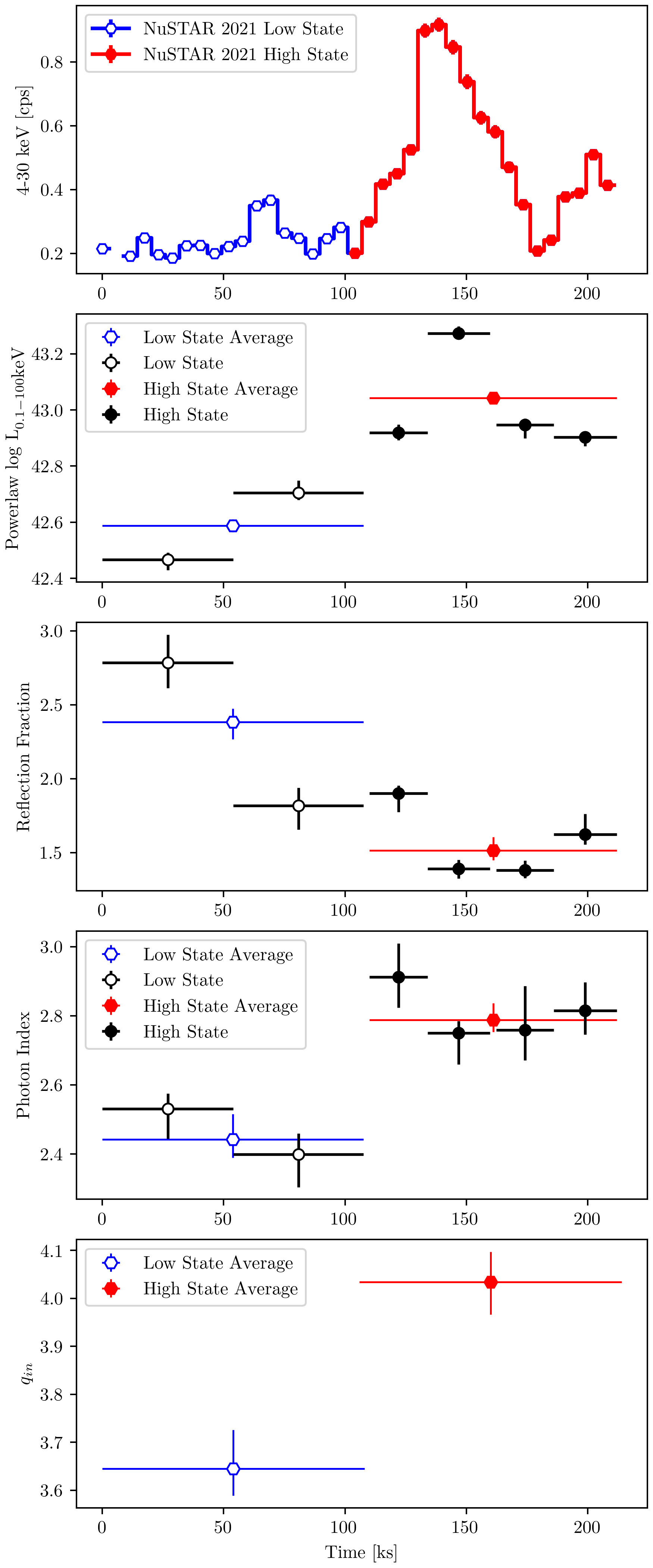}
    \caption{
    \textit{Top panel:} \nustar\ $4-10$ \kev\ light curve. \textit{Second panel:} The log $0.1-100$ \kev\ luminosity of the power law component. \textit{Third panel:} Reflection fraction calculated using the observed $0.1-100$ \kev\ luminosity of each component.  \textit{Forth panel:} The photon index, and \textit{Bottom panel:} The inner emissivity index for the average low and high states.}
\label{fig:index1} 
\end{figure}

To investigate the flaring event caught with \nustar, we subdivided the low and high-flux intervals into two and four spectra, respectively (see Figure \ref{fig:index1}).  Each subdivided spectra contained $2400-6600$ counts summed between FPMA and FPMB. They averaged $\sim25.6\ks$  and $\sim12.4\ks$ total exposure time ($\sim50\ks$ and $\sim25\ks$ duration) for the low and high states, respectively. These spectra were fit along with the average high, low and PN spectra using the parameters found from the best-fit model. For each spectrum, we allowed only the photon index, reflection fraction, and power law normalization to vary.  We report these parameters in Figure \ref{fig:index1}. The empty blue hexagons indicate the average low state value, and the filled red hexagons show the average high state value. The empty black circles show the subdivided low-state values and the filled black circles show the subdivided high-state values.

The luminosity of the power law component increases for the first 150\ks\ before dropping slightly after the flare peaks. The reflection fraction in the low state decreases as the power law luminosity increases. During the first 50\ks, the system is in a true low state, as we saw in 2019. Then, between 50 and 100\ks, the corona exhibits a small flare and increases in brightness. Correspondingly, the reflection fraction is reduced. 

Between 100 and 125\ks, the main flare begins, and the power law luminosity increases again. However, the reflection fraction initially remains constant, indicating that emission from the disc has increased. The flare peaks just before 150\ks. This corresponds to the largest power law luminosity and the lowest reflection fraction. The reflection fraction remains low as the power law luminosity begins to fall off after the flare before increasing slightly by the end of the \nustar\ observation.

The reflection fraction is not obviously inversely correlated with the power law on the short time scales. This would be the behaviour expected if the blurred reflection component was constant on rapid time scales; changes in the power law luminosity would be the only component driving the changes in reflection fraction. However, this is not what is observed. At some point, photons emitted by the corona reflect off the disc and then are remitted out of the system. 
This might indicate some delay between the power law and reflection components, which is an expected consequence of blurred reflection \citep{Fabian+2009B, Zoghbi+2010, Wilkins+2014}. Unfortunately, we do not have the data quality to probe on finer time scales. 

The rapid flare seen over the $4-30$ \kev\ band points to a significant level of intrinsic AGN variability in \mrk\ that is comparable to other NLS1s. Figure \ref{fig:index1} reveals several observed flux-related parameter changes. We can see that the change between the low and high states happened quickly for the photon index and inner emissivity index.
This change in $\Gamma$ between the low and high states suggests that the physical conditions in the corona change during the \nustar\ observation. The photon index of a corona is determined by its Compton-$y$ parameter \citep[see:][]{Longair+1992, Rybicki+1979, Fabian+2015}, which is dependant on the temperature ($T_e$) and the optical depth ($\tau_e$) of the emitting plasma. Presumably, either or both of these parameters may have changed during the flare. It is difficult to isolate the interesting parameters as they are degenerate. Measurement of the cut-off energy would be beneficial \cite[e.g.][]{Wilkins+2014, Wilkins+2022}, but this could not be constrained with the current data. 

The change in the inner emissivity index is the opposite of what we might expect from a corona with lamp-post geometry. In a lamp-post geometry, the corona is considered a point source at some height $h$, above the accretion disc. If the height of the corona changes by moving to a larger $h$, then the number of continuum photons escaping the system would increase, leading to an increase in flux and a decrease in reflection fraction, as we see in \mrk. However, this increase in height would lower the inner emissivity index as fewer continuum photons would be incident on the inner disc \citep{Wilkins+2012}.  Furthermore, under a lamp-post geometry, the break radius we observe suggests a coronal height of $\sim50$\rg\ \citep{Wilkins+2012}. At this height, we would not expect to see the high reflection fraction we observe in this AGN \citep{Gonzalez+2017}.

The increase in photon index, decrease in reflection fraction and increase in the inner emissivity index mirror the behaviour of Mrk\,335 during a proposed jet launching event \citep{Wilkins+2015B}. The changes in the parameters we see in \mrk\ are far less extreme but do follow the same trend. This could indicate a vertically extended corona, which would produce an emissivity profile very similar to a broken power law, with the inner emissivity index being slightly steeper than the outer index \citep{Wilkins+2012}. That is consistent with the emissivity profile we measured in \mrk. 

\section{Discussion}
\label{sec:disc}
\subsection{The Origin of the Soft X-ray Emission Components }
\label{sec:ISM}
Unlike most NLS1s, the central engine of \mrk\ is hosted in an elliptical \citep{Konig+2009} or compact galaxy\citep{Mazzarella+1986}. It has been classified as unbarred and bulge-dominated \citep{Husemann+2022}. It is known that the hot gas in such galaxies can produce emission detectable in the soft band related to their hot interstellar medium (ISM) \citep{Fabbiano+2012}. In particular, there are well-known relations between the X-ray luminosity and the optical and infrared bands. Here, we compare the soft band luminosity of \mrk\ to that of other S0 and elliptical galaxies. 

The luminosity of the ISM in the 300\kpc\ surrounding the massive elliptical galaxy, NGC\,4636, is $L_{0.5-4.5{\rm  \kev}}=10^{41.9}$\ergps\ \citep{Matsushita+1998}. NGC\,4636 has a total mass of $9\times 10^{12}$\Msun\ \citep{Matsushita+1998}, and a stellar mass of $7\times10^{11}$\Msun\ \citep{Mathews+2021}. The X-ray emission of the ISM is spatially resolved. Its RGS spectra can be fit with an {\sc apec} component whose temperature increases with radius from 0.53 \kev\ to 0.71 \kev\ \citep{Xu+2002}. The temperatures observed in NGC\,4636 are comparable to what we observe in \mrk. The $0.5-4.5$ \kev\ luminosity of just the soft band components in \mrk\ is $L_{0.5-4.5{\rm  \kev}}=10^{41.7}$\ergps, 0.2 dex less than NGC\,4636. However, \mrk\ is relatively isolated and has a much smaller stellar-mass  ($M_*=1.5\times10^9$\Msun\ \citealt{Pan+2019}; $M_*=1.0\times10^{10}$\Msun\ \citealt{Smirnova+2022}) as compared to NGC\,4636 \citep[$M_*\sim10^{12}$\Msun][]{Mathews+2021}.

The $L_X-L_B$ relation found in ellipticals shows that at higher blue luminosities, the X-ray luminosity also increases \citep{Diehl+2007}.  To compare \mrk\ to the sample of elliptical galaxies, we first estimate the B-band luminosity to be $L_B=10^{10}L_{\odot, B}$ based on an SDSS $u$ magnitude $u=16$ and a luminosity distance of $D_L=86$\,Mpc \citep{Sani+2010}. The absorbed X-ray luminosity over $0.3-5$ \kev\ of the collisional and constant continuum components combined  were $L_{X,T}=10^{41.7}$\ergps\ ($L_{X,{\rm CIE}}\sim10^{41.2}$, $L_{X,{\rm CC}}=10^{41.5}$\ergps,  for collisional and constant continuum components, respectively). Comparing to figure 4 of \cite{Diehl+2007}, we see that the measured $L_B$ indicates that the luminosity of the soft band components in \mrk\ should be of order $L_{X}=10^{40}-10^{41}$\ergps. The observed values are greater than expected. If we explore the unabsorbed luminosity the discrepancy increases. The unabsorbed luminosity of the collisional component becomes  $L_{X,{\rm CIE}}\sim10^{41.8}$\ergps, while the unabsorbed luminosity of the constant continuum component becomes $L_{X,{\rm CC}}=10^{42.7}$\ergps.The total luminosity then becomes $L_{X,T}=10^{42.8}$\ergps.

Similar relationships exist between the stellar luminosity, as measured by the $K$-band, and the X-ray luminosity of the ISM of elliptical galaxies \citep{Boroson+2011}. Repeating the above exercise with $K$ magnitudes from 2MASS, we take $K=9.6$, resulting in $L_K=10^{11.4}L_{\odot, K}$. Comparing this to figure 2 of \cite{Boroson+2011}, we see that the luminosity of each X-ray component and the total luminosity of \mrk\ is greater than the hot gases, low mass X-ray binaries (LMXB), and nuclear emission in all of the elliptical galaxies probed. 

We note that the $B$ and $K$-band luminosities of \mrk\ have not been corrected for AGN emission, and thus, the discrepancy between the expected X-ray luminosity and what we observe is only a lower limit here. As such, the discrepancy here indicates that the soft band emission we see in \mrk\ does not originate purely from hot gases often seen in elliptical galaxies. Instead, this emission must be associated with some process indirectly powered by the AGN. 

The soft X-ray emission in \mrk\ appears to originate from three distinct components.  Based on the \chandra\ image of \mrk, we know that the soft X-ray emission originates within 500\pc\ of the central engine and is asymmetric \citep{Buhariwalla+2023}. 

These soft band components may be common in all AGN  with radio jets and are only visible in \mrk\ due to the heavy absorption present. The unabsorbed AGN continuum luminosity between $0.2-2$ \kev\ is log$L_{0.2-2 \kev}=43.2$\ergps. The luminosity of the bremsstrahlung component in the same band (log$L_{0.2-2 \kev}=41.8$ \ergps) is less than 4 percent of the AGN luminosity. All three soft band components would represent $\sim8$ percent of the AGN luminosity in the soft band. A simulated spectrum with no continuum absorption and the three soft band components can be fit with just the continuum contributions from {\sc relxill} and {\sc po} (see Appendix \ref{B:PN}). This indicates that these components could very well be present in other galaxies and simply be overwhelmed by the central engine. We can see this in the top panel of Figure \ref{fig:pl_eem_ld_ra_low}, where the unabsorbed model is $\sim1$ dex above most emission lines and $\sim2$ dex above the soft band continuum. 

\subsubsection{The Photoionized Emission} 
\label{sec:PIE}
The photoionized component was modelled with {\sc xillver} in the CCD spectra, whereas the photoionized component in the RGS spectra was modelled with an {\sc xstar} grid \citep{Buhariwalla+2023}. The reason why {\sc xillver} is a poor fit in the RGS spectra is that it is at a sufficiently high density ($10^{15}/{\rm cm}^3$), the forbidden line of the \ovii\ triplet is suppressed. In the PN spectra, we cannot resolve any individual line in the soft band, thus  {\sc xillver} appears adequate. 

If we very closely examine the ratio of models 7 and 6b ({\sc zpcfabs*apec+zbrem+xstar}/{\sc xillver}, for 7/6b, respectively), the discrepancy between the models lies mostly between $0.4-0.6$ \kev. This is the location of the two most prominent photoionized features found in the RGS spectrum, the \ovii\ triplet and \nvii\ line. While {\sc xillver} does not reproduce the \ovii\ triplet expected in this energy band, it is at such low resolution that all three \ovii\ lines blend together in addition to the \nvii\ line. We note that \cite{Buhariwalla+2023} were underfitting the \nvii\ line at 25\AA\ (0.495 \kev) with their best-fit model. Thus, {\sc xillver} is better at accounting for the emission in this narrow band. The true nature of the photoionized material thus must lie somewhere in between the {\sc xstar} and {\sc xillver} models. 

The physical location of the photoionized material is at some distance greater than the dust sublimation radius, and it is outflowing at a velocity of $\sim660$ \kms\ \citep{Buhariwalla+2023}. We see no evidence of absorption of this component, indicating that we have a clear line-of-sight to the site of the photoionized emitter. The photoionized emitter could be originating from an outflow on torus scales. We note that the photoionized emitter is also responsible for fitting narrow features in the \feka\ band. This assumes that the region responsible for producing the \feka\ is the same as producing the \ovii\ triplet emission.

Understanding the continuum photoionizing the gas is imperative to determine the location and nature of the photoionized emission. With the RGS data, we were unable to distinguish different illuminating sources. However, the PN data can offer some insight. There is no fit improvement by allowing the {\sc xillver} component to have $\Gamma$ free and distinct from the power law continuum. Also, fixing the photon index of the photoionizing continuum to the value of the power law continuum is indistinguishable from fixing at the fiducial value of $\Gamma=2.5$. We can, however, constrain the photoionizing continuum photon index to be $\Gamma>2.2$. This indicates that the primary X-ray continuum is responsible for ionizing the photoionized emission, and thus, there must be a relatively clear line of sight between the corona and the photoionized material. Furthermore, there must be a clear line of sight between the photoionized material and the observer. To accomplish this, the neutral absorber must be asymmetric. Details of the geometry of the system will be discussed further in Section \ref{sec:Geo}. 

\subsubsection{The Collisonally Ionized Emission}
Based on the best-fit model, the CIE ({\sc apec}) component is absorbed. This results in an intrinsically brighter {\sc apec} component than was previously reported in \cite{Buhariwalla+2020}. The new higher value of $L_{2-10 \kev}$ results in a new SFR estimate  between $25-35\Msun{\rm yr}^{-1}$ \citep{Franceschini+2003}. Based on $L_{0.5-2 \kev}$, the SFR would be in excess of $100\Msun{\rm yr}^{-1}$ \citep{Ranalli+2003}. 

It becomes much less likely that this collisionally ionized emission is originating in regions of star formation as this is orders of magnitude greater than the SFR predicted in other band passes [PAH measurements, $\leq$ 7.5\Msun\,yr$^{-1}$, \citep{RuschelDutra+2017}; SED fitting, $3.47\pm 0.26$\Msun\,yr$^{-1}$ \citep{Gruppioni+2016}; IR measurement, $2.1_{-0.4}^{+0.5}$\Msun\,yr$^{-1}$, \citep{Smirnova+2022}]. 

An alternative could be the shock heating of material as it is pushed into the ISM by the jet or other outflow mechanisms present in this galaxy. Modelling the soft band with a plane parallel shocked plasma model, {\sc pshock}, produces a fit of equal quality as with {\sc apec} ($\Delta{\rm AIC}=5$, favouring {\sc apec}).

\subsubsection{The Featureless Continuum}
\label{featurlessCont}
There is a constantfeatureless continuum component evident in the soft band. In \cite{Buhariwalla+2020}, this component was attributed to the leaky absorber permitting AGN continuum from the central engine to contaminate the soft band. Our timing analysis has demonstrated the AGN varies significantly, and any leaked emission should also vary. However, the soft band does not exhibit such behaviour. Instead, we can fit the featureless continuum using either a bremsstrahlung ({\sc zbrem}) or a black body ({\sc zbb}) component, with the bremsstrahlung component being preferred statistically. This component was not detected in the RGS spectra. The flux between a $0.5-1.77$ \kev\ ($7.3-24.7$\AA) of the {\sc zbrem} component in the PN spectra is $\sim1.2\times10^{-13}$ \ergpscmps. The flux of the RGS background in this region is $\sim6\times10^{-13}$ \ergpscmps, or approximately five times brighter. Thus, the null detection of the constant continuum component in the RGS was likely a sensitivity issue. 

It is unlikely that the origin of the constant continuum component lies with an actual bremsstrahlung emitting plasma.  Firstly, the model component, {\sc zbrem}, relies on a thermal distribution of electrons \citep{Kellogg+1975}, while the most probable location for a bremsstrahlung emitting plasma to exist in an AGN is inside a radio jet. Those electrons, however, would have a non-thermal distribution and thus would produce a non-thermal spectrum.

If we consider it a thermal bremsstrahlung plasma, then we find this is also not plausible. The temperature of the {\sc zbrem} component of kT $\sim0.2$ \kev. The dominant continuum process of plasma at that temperature would not be bremsstrahlung radiation, but rather radiative recombination and/or 2-photon continuum \citep{Bohringer+2010}. These are the dominant continuum processes in {\sc apec} at this low temperature. We attempted to fit the soft band with two collisionally ionized components. The problem with this scenario is that at low temperatures, emission lines dominate over all continuum processes \citep{Bohringer+2010}. Thus, to explain the presence of the constant continuum component using only collisional plasma, we would require the abundance of that plasma to be primordial, as this would effectively remove all emission lines. This is allowed by the data but stretches the imagination as to how such a plasma could exist.

Another option is that the origin is thermal emission from warm material. The temperature of the black body is ${\rm kT}\sim90\,{\rm eV}$ (${\rm T}\sim10^6\,{\rm K}$). This is an order of magnitude hotter than we would expect for the accretion disc of an SMBH \citep{Hickox+2018}. Even if \mrk\ was host to an exceptionally hot accretion disc, we would not expect to see it directly as it would also be obscured through the $\nh=10^{23.5}\pscm$\ absorber and thus not visible. 

A final possible explanation for the excess emission is calibration issues between the RGS and PN instruments. There is a known correction needed between these instruments \citep[See appendix B of ][for details]{Grafton+2023}. However, we attempted to mitigate these effects by ignoring PN data below 0.5\,\kev, and we found that the excess emission was still present. 

It is difficult to say anything more based on these data and deeper observations of \mrk\ with instruments like \textit{Arcus} \citep{Smith+2019} and \textit{Athena} \citep{Nandra+2013} to explore the soft band more fully. 

\subsection{The Geometry of Mrk 1239}
\label{sec:Geo}
In this work, we estimate the inclination of the accretion disc to be $\sim17^{\circ}$, based on the blurred reflection model. The inclination of the torus was measured to be $\sim6^{\circ}$ \citep{Lakicevic+2022}, which was consistent with the measurements of inclination of the broad line region ($7^{\circ}$, \citealt{Zhang+2002}; $11^{\circ}$, \citealt{Gravity+2024}). We conclude that we are observing the system virtually face-on. The neutral absorber is unlikely to be directly associated with the torus at this inclination, as it would have to be nearly spherical. The opening angle of the broad line region was measured to be $\sim42^{\circ}$, by \cite{Gravity+2024}. If we assume that the torus is a natural evolution of the BLR \citep{Koshida+2014, Minezaki+2019, Gallo+2023} then the opening angle of the torus must be similar to $\sim42^{\circ}$.

Two polarization regions are present in \mrk, one associated with the dust in the broad line region \citep{Goodrich+1989}, the other associated with polar dust \citep{Batcheldor+2011, Lakicevic+2022}. The polar dust region is responsible for the bulk of the polarization seen in the optical spectrum of \mrk. This dust could be associated with the dust causing the attenuation. Furthermore, it could be associated with the IR bump due to hot dust that is present in \mrk\ \citep{Rod+2006}. This dust, however, is not hot enough to contribute to the featureless continuum we see in \mrk.  

\cite{Rod+2006} measure an extinction of $E(B-V)=0.54$, in the inner few $\sim100\pc$ of the central engine in \mrk. Based on the dust-to-gas ratio of obscured quasars \citep{Jun+2021}, this results in a column density of $1-4\times10^{22}$\pcm.
The obscurer measured by \cite{Pan+2021} close to the central engine had an extinction of $E(B-V)=1.6\pm0.1$, which resulted in a column density between $4\times10^{22}$ and $1\times10^{23}$ \pcm. This is closer to the column density we measure in the X-ray regime ($\sim3\times10^{23}$\pcm). The discrepancy could be due to uncertainty in the dust-to-gas ratio used to infer the column density of the optical obscurer.

Assuming a BH mass of $2.6\times10^6$\Msun\ \citep{Marin+2016, Gravity+2023}, the Eddington luminosity is $33\times10^{43}$\ergps. During the low state, the unabsorbed $0.1-100$ \kev\ luminosity of the central engine is estimated to be $\sim7 \times 10^{43}$\ergps. Using this as a proxy for the bolometric luminosity gives an Eddington ratio of $\lambda=0.2$. This is consistent with the value found by \cite{Jiang+2021}. Based on these values, we expect the absorbing material in \mrk\ to fall within the forbidden region of the $\nh-\lambda$ plane \citep{Ishibashi+2018}. Thus we would expect the neutral absorbing material to be outflowing, which has been observed at $\sim1000\,$\kms\  \citep{Pan+2021}. Comparing the absorbed and unabsorbed luminosity of the central engine, we find that $\sim6\times10^{43}$ \ergps\ are absorbed by the neutral material. 

 It seems this neutral absorber is asymmetric, extending much further on one side of the central engine than the other. Assuming this neutral material is what is crashing into the ISM, creating the collisionally ionized and accountant continuum emission, then this asymmetric cloud is consistent with the asymmetric soft emission that we see in the \chandra\ image of \mrk\ \citep{Buhariwalla+2023}.

\begin{figure}
    \centering
    \includegraphics[width=\linewidth]{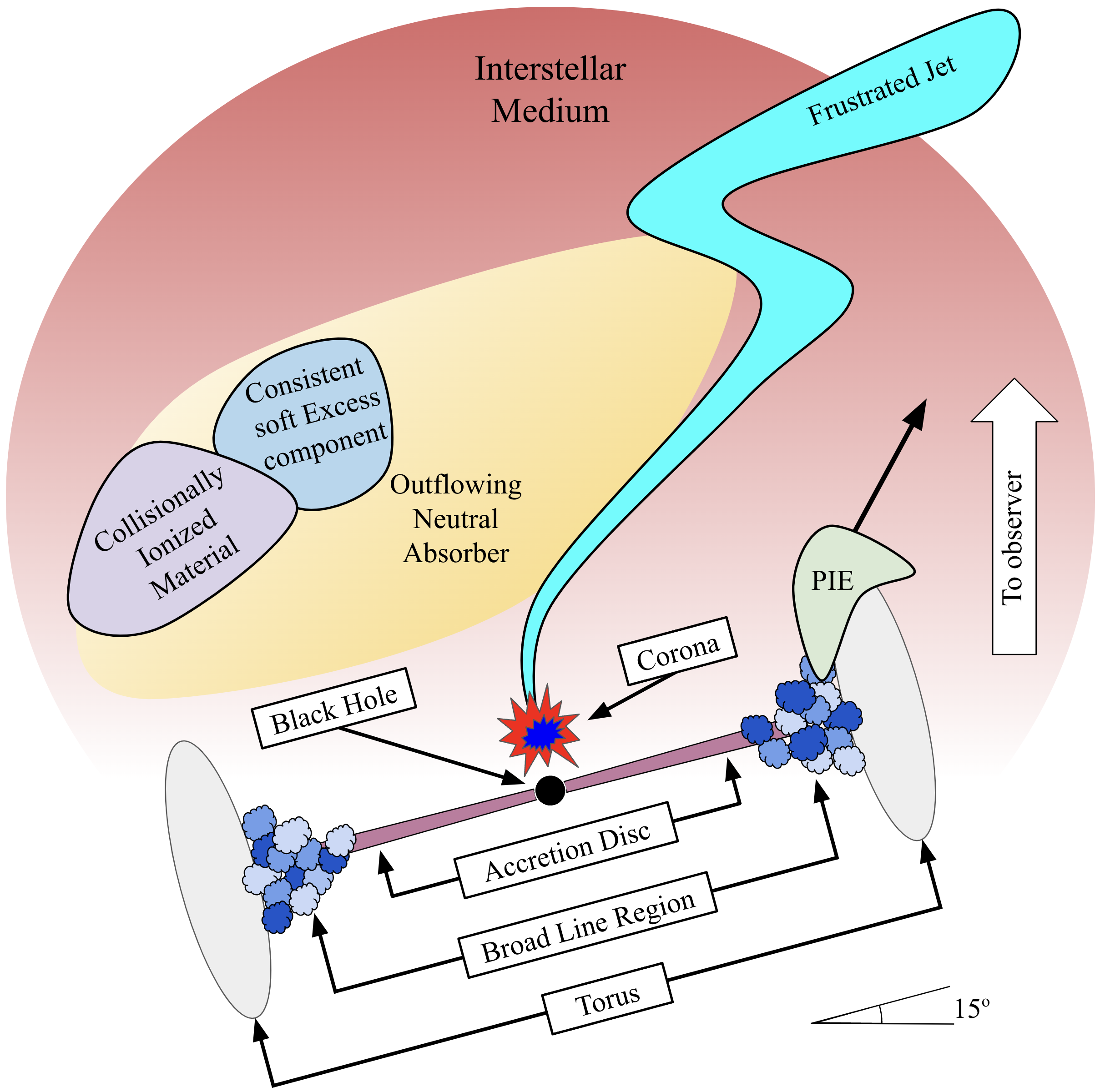}
    \caption{A possible profile view of the central engine in \mrk. The standard components of an AGN are labelled, e.g. torus, broad line region, accretion disc, black hole and corona. The larger red corona is in the high state, while the smaller blue corona represents the low state. The corona is the base of a vertically extended jet (cyan) that pushes material to the neutral absorber.  This absorber represents the $10^{23}{\rm cm}^{-2}$ absorber that obscures the continuum emission from \mrk. The outflowing neutral absorber crashes into the ISM (red), producing the collisionally ionized and possibly the constant continuum component. The photoionized material (PIE) is located on the other side of the central engine, where there may be a clear line of sight at torus-like distances.}
    \label{fig:profile_view}
\end{figure}

The outflowing velocity of this cloud is comparable to the velocity measured from the forbidden \ovii\ line \citep{Buhariwalla+2023}, suggesting a possible common origin. However, as discussed in Section \ref{sec:PIE}, the photoionized emitter must have a clear line of sight to both the central engine and the observer, suggesting they are physically distinct. If we assume that the photoionized absorber is being blown off of the top of the torus at $\sim45^{\circ}$ \citep{Gravity+2024}, then the resultant rest frame velocity of the photoionizing material would be $\sim1000$ \kms.  The velocity would be different depending on the location on the torus that the PIE originates from. If the emitting region was tilled towards us as seen in Figure \ref{fig:profile_view}, then the rest frame outflow velocity might be closer to  $\sim800$ \kms. If the PIE was located on the exact opposite side of the tours, tilling away from us,  then the outflow velocity might be closer to $\sim1300$ \kms. 

Compact steep spectrum radio sources (CSS) may be young radio-loud galaxies trying to push their way out of the central region \citep{Fanti+1990, ODea+1998}. The outflow velocity of these objects is typically $\sim2000\,$\kms\ \citep{Fanti+1990}. In the radio, \mrk\ has a steep spectral slope \citep{Berton+2018, Jarvela+2022} and it is incredibly compact. About 90\% of the radio emission is contained within the central $\sim10$s of parsec \citep{Doi+2015}. However, the jets are subsonic and not well collimated and instead might be considered a low-luminosity compact source whose radio jets have become disrupted or frustrated by the ISM and other material surrounding the central engine \citep{Kunert+2010, Doi+2015, Jarvela+2022}. 

The kinetic energy of the jet is estimated to be $\sim10^{43}$ \ergps\ \citep{Doi+2015}. The bolometric luminosity of the collisional material (collisionally ionized and bremsstrahlung)  is $10^{43}$ \ergps\, which implies that $\sim100\%$ of the jet kinetic energy is required to produce the collisional luminosity that we see in \mrk. While the jet may provide some kinetic energy to the outflowing material, it is unlikly to be the sole contributor.  

Figure \ref{fig:profile_view} shows our proposed central engine geometry for \mrk. The system is viewed virtually face-on (fiducial inclination of $15^{\circ}$)  and contains all the standard components for an AGN (corona, accretion disc, BLR, tours). Where \mrk\ differs from the typical Type I galaxy is the massive asymmetric neutral absorber that is crashing into the ISM. The radio jet or radiatively driven outflow (or both) pushes the neutral material out of the system. This obscuring cloud is larger than the BLR and torus and covers a significant fraction of both. A region at the approximate distance to the torus remains unobscured; this is where the photoionized emission originates from. The exact nature and location of the consistent continuum component are unknown; we have placed it near the CIE because the data allows (but does not require) both components to be absorbed by the same neutral material. 

Based on the work of \cite{Ishibashi+2018}, we expect the absorbing material to expand (blow-out) such that the continuum starts to overwhelm any collisionally ionized material that may be present within a few hundred thousand years. The criteria for this is the time it takes for an absorber with a column density of log$\nh=24$\pscm\ to expand and reach log$\nh=22.5-22.75$\pscm. At these column densities, we would not expect to see the peculiar soft band behaviour that \mrk\ exhibits. If the AGN started with a lower column density, or if radiation pressure were able to preferentially blow out a section of the absorber such that the total covering fraction was reduced, then this time may be shorter. We expect the AGN duty cycle to last $\sim10^7-10^8$ years \citep{Parma+2007}. Assuming a duty cycle of $50\,$Myr and a blow-out time of $100\,$kyr, then it would take $\sim 0.2$ percent of an AGN active time to blow an absorber out such that it no longer displayed the collisionally ionized plasma features like \mrk\ possesses. Perhaps \mrk\ is preferentially blowing the material from one side, and this is what is creating the asymmetric nature of the absorbing cloud (see Figure \ref{fig:profile_view}). If the absorbing cloud has been preferentially removed at one side of the tours, then this could also be a possible explanation as to why the photoionized emitter is outflow.

This indicates that observing a type 1 Seyfert galaxy with this level of absorption is a rather rare event. Of the more than 800 Seyfert galaxies in the \swift-105 month survey \citep{BAT+105+month+survay}, perhaps two share the unique timing and spectral properties that \mrk\ exhibits. The challenge of identifying these objects is that it is not just that they are short-lived and few in number but that they require careful timing and spectral analysis to identify. Sample searches in the \swift-LXSPS catalogue \citep{Evans+2023}  and \textit{eROSITA} \citep{Predehl+2021, Merloni+2024} may be able to identify more of these objects. 

\section{Conclusions}
\label{sec:conclusion}
We have revealed the behaviour of the central engine in new detail with the broadband long-term monitoring of \mrk. We perform timing analysis on long-term (\swift\ data) and rapid (PN and \nustar\ data) time scales. We performed broadband X-ray spectral modelling with flux-resolved CCD spectra and compared our results to archival data of \mrk, finding good agreement. 
\begin{itemize}
  \item The timing analysis of \mrk, including the two \swift\ monitoring campaigns, shows that the central engine of \mrk\ is totally obscured by an $\nh\sim10^{23}$\,cm$^{-2}$ absorber. No central engine continuum variations are detected below 3 \kev.  Instead, the soft band emission originates on a larger physical scale. It is comprised of three main components: a collisionally ionized, photoionized, and a constant continuum component shaped similar to bremsstrahlung emission. The collisionally ionized and constant continuum components could be associated with a large-scale outflow crashing into the ISM, While the photoionized component is most likely related to the material at tours-like distances that have been photoionized by the unabsorbed primary continuum of \mrk.
  \item The collisionally ionized plasma requires some absorption. Consequently, a star forming region is no longer favoured because it would require very high rates inconsistent with other measurements. Instead, we conclude that this region is consistent with outflowing material crashing into the ISM. 
  \item The large amplitude, rapid flaring event seen in the data up to 30 \kev can be attributed to a brightening of the primary continuum and not due to obscuration. In combination with the presence of a Compton hump, a blurred reflection origin for the primary continuum is favoured.  
  \item The accretion disc of \mrk\ is viewed virtually face-on, and it produces a strong reflection spectrum. The flaring event observed in \mrk\ was due to changes in the corona that propagated to changes in the accretion disc. The measured emissivity profile is consistent with a vertically extended primary source at large distance, possibly indicating that the base of the jet is serving as the corona.   
\end{itemize}
We estimate less than 1 percent of AGN may have geometries similar to \mrk. However, \mrk\ might be revealing a stage of rapid evolution in all AGN when the central region is being cleared of dense material and exposing the central engine. Further study is required to examine this type of galaxy.

\begin{acknowledgments}
We thank the referee for their helpful comments that improved this manuscript. This work was based on observations obtained with XMM-Newton, an ESA science mission with instruments and contributions directly funded by ESA Member States and NASA. LCG acknowledges financial support from the Natural Sciences and Engineering Research Council of Canada (NSERC) and the Canadian Space Agency (CSA). JJ acknowledges support from the Leverhulme Trust, the Isaac Newton Trust, and St. Edmund's College University of Cambridge. 
\end{acknowledgments}
\section*{Data Availability}
All data are publicly available through the HEASARC archive.


\bibliographystyle{aasjournal}
\bibliography{bibtext} 


\appendix
\section{\swift\ spectral Data}
\label{A:swift}
\begin{figure}
	\centering
	\includegraphics[width=0.8\linewidth]{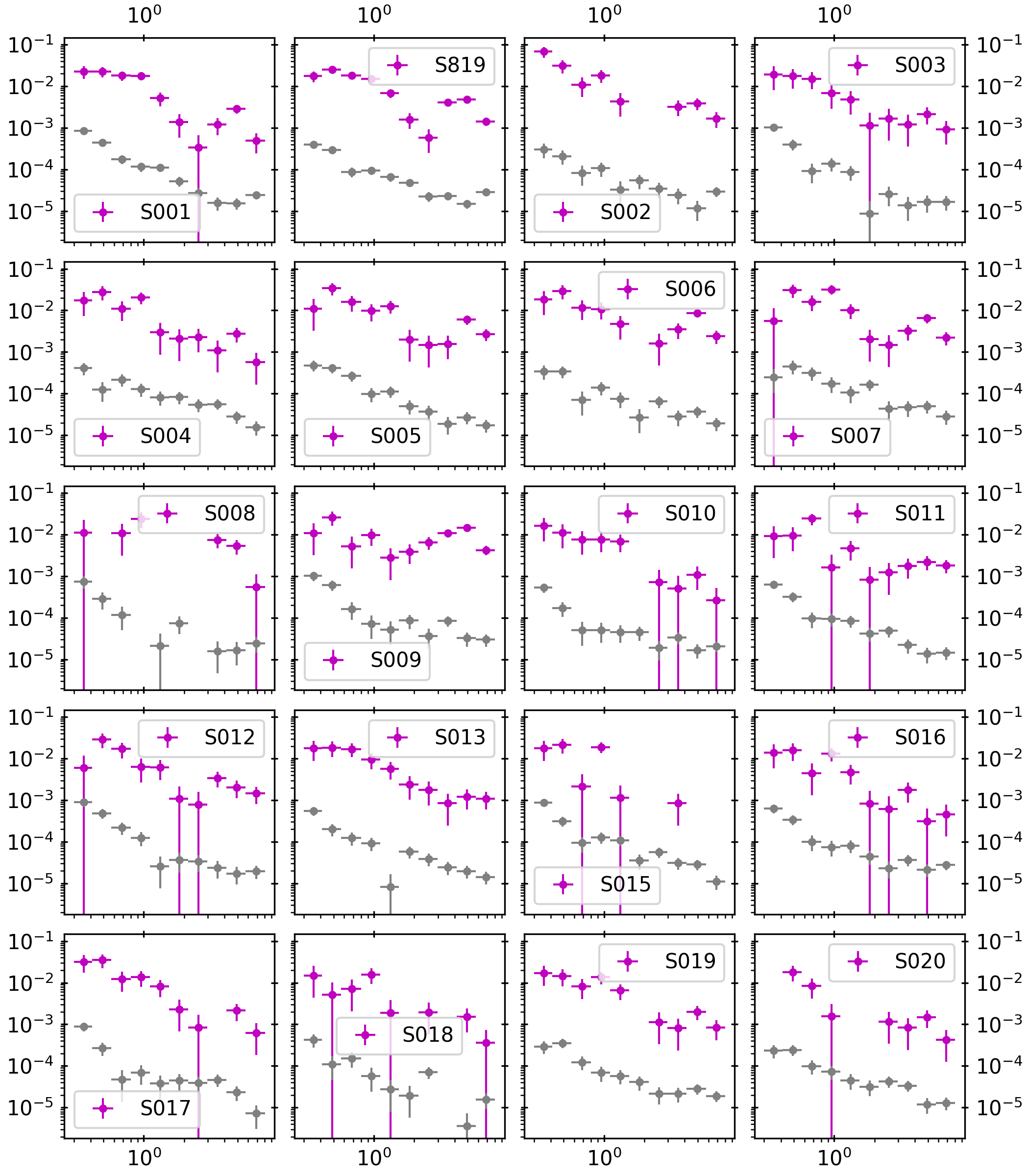}
	\caption{\swift\ spectrum for observations detailed in Table \ref{tab:SwiftObs}. The data has been binned into 10 bins to match the first principal component seen in Figure \ref{fig:PCA}. The gray data represent the background for each \swift\ observation.}
	\label{fig:July12_swift5x4.png}
\end{figure}
\section{Simulated PCA Principal Component 1}
\label{D:pca}
\begin{figure}
	\centering
	\includegraphics[width=0.8\linewidth]{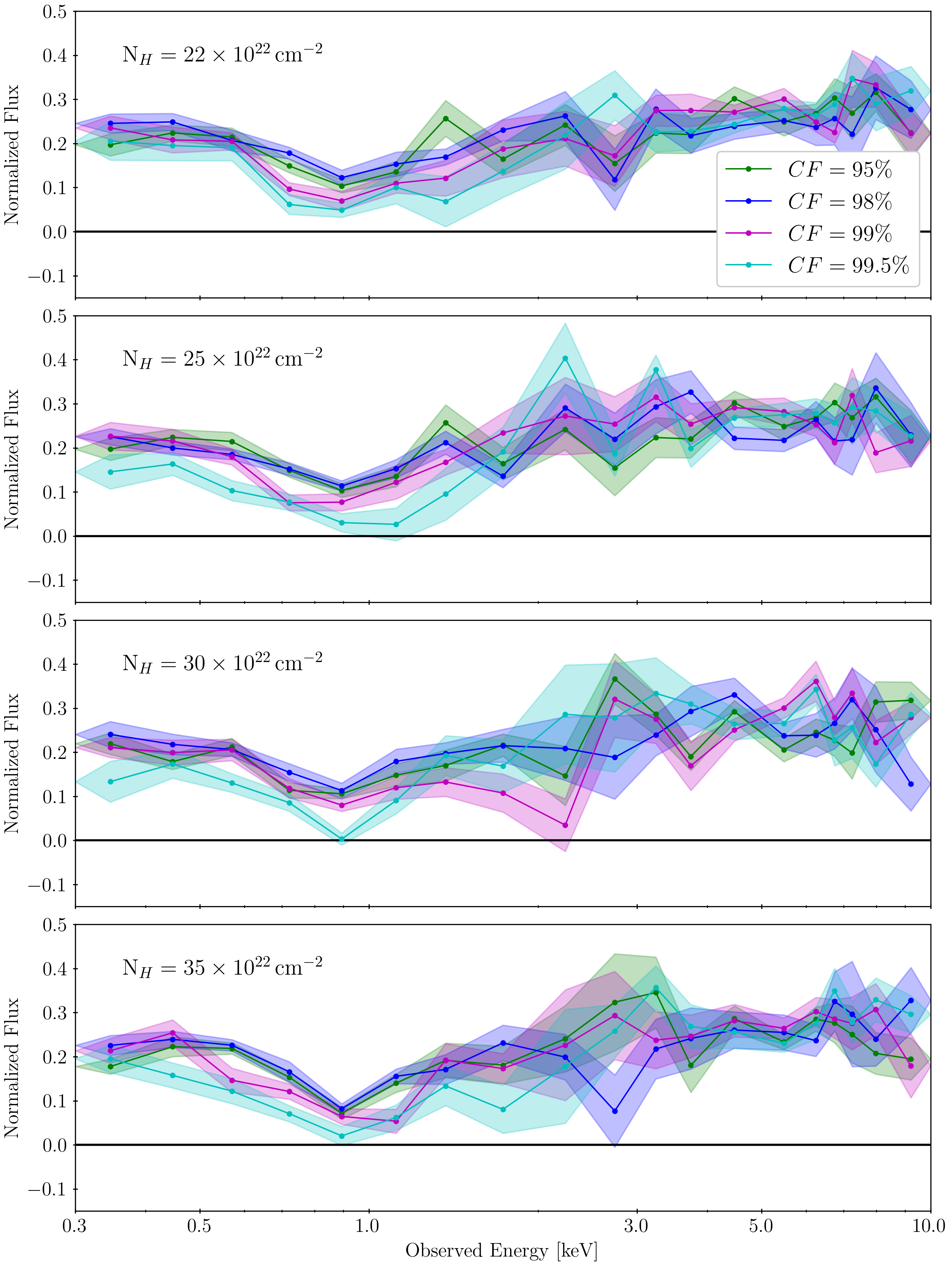}
	\caption{The first principal component for the sixteen simulated absorbers described in Section \ref{sec:TIMEconc}. The absorbers are grouped by column density, with the top panel showing those with $\nh=22\times10^{22}$ \pscm, the second panel showing those with  $\nh=25\times10^{22}$ \pscm, the third panel showing those with $\nh=30\times10^{22}$ \pscm, and the forth panel showing those with $\nh=35\times10^{22}$ \pscm. Each panel has absorbers with covering fractions of $CF=95,\,98,\,99,$ and $99.5\%$. All simulated absorbers produced a first principal component that was inconsistent with the first principal component of the PN spectra.} 
	\label{fig:april30x2024_PCA_sims.png}
\end{figure}
\section{Simulated PN data}
\label{B:PN}
\begin{figure}
	\centering
	\includegraphics[width=0.5\linewidth]{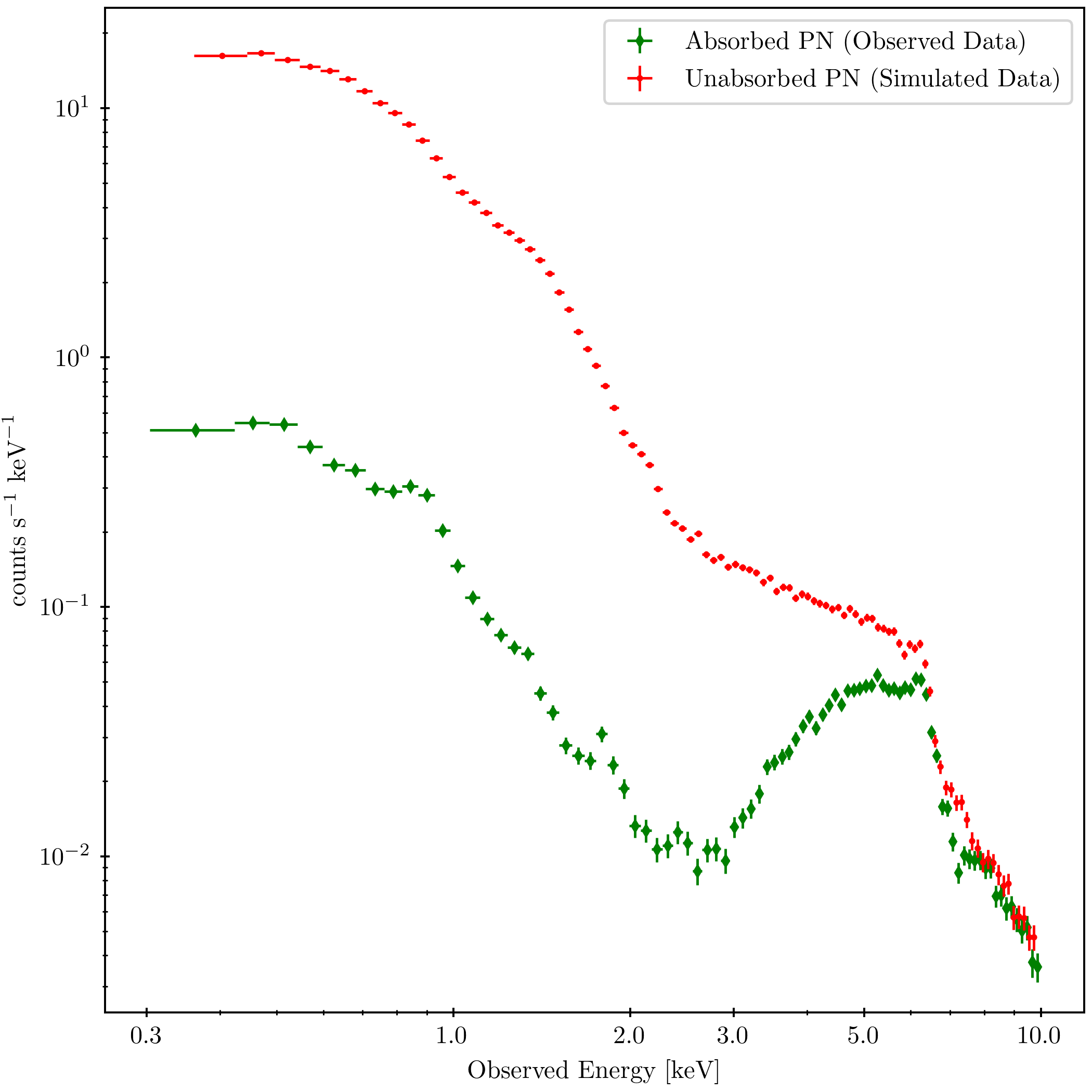}
	\caption{The observed PN spectra (green diamonds) compared to a simulated PN spectra based on the best fit low-state model with no absorption. See section \ref{sec:lowStateSpectra} for details. }
	\label{fig:realPNfakePN}
\end{figure}

\section{MCMC corner plots}
\label{C:mcmc}
\begin{figure*}
	\centering
	\includegraphics[width=\linewidth]{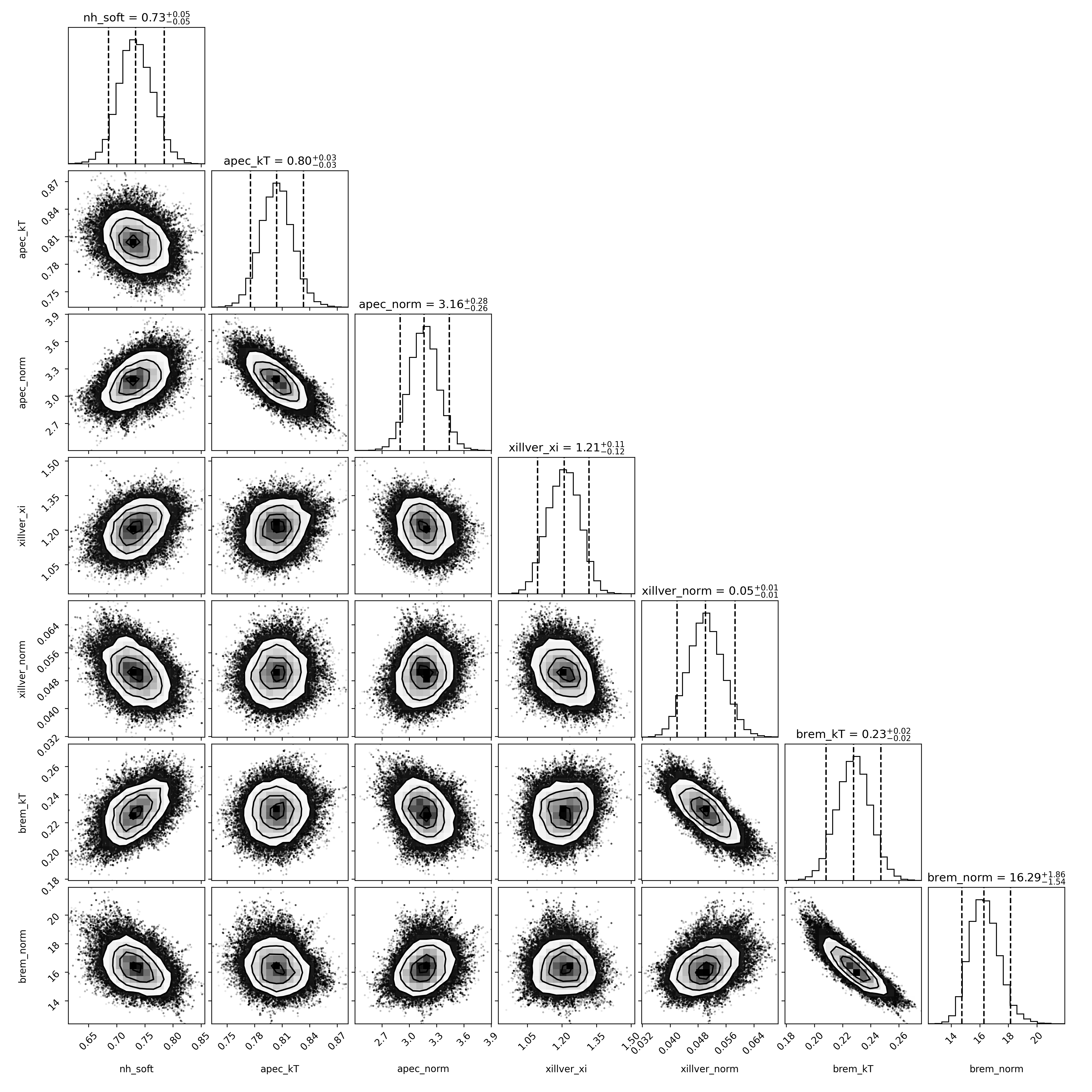}
	\caption{The corner plots for the MCMC best fit for the soft band components. Note that nh\_soft is the column density of the absorber applied to the {\sc apec} component.}  
	\label{fig:softMCMC}
\end{figure*}
\begin{figure*}
	\centering
	\includegraphics[width=\linewidth]{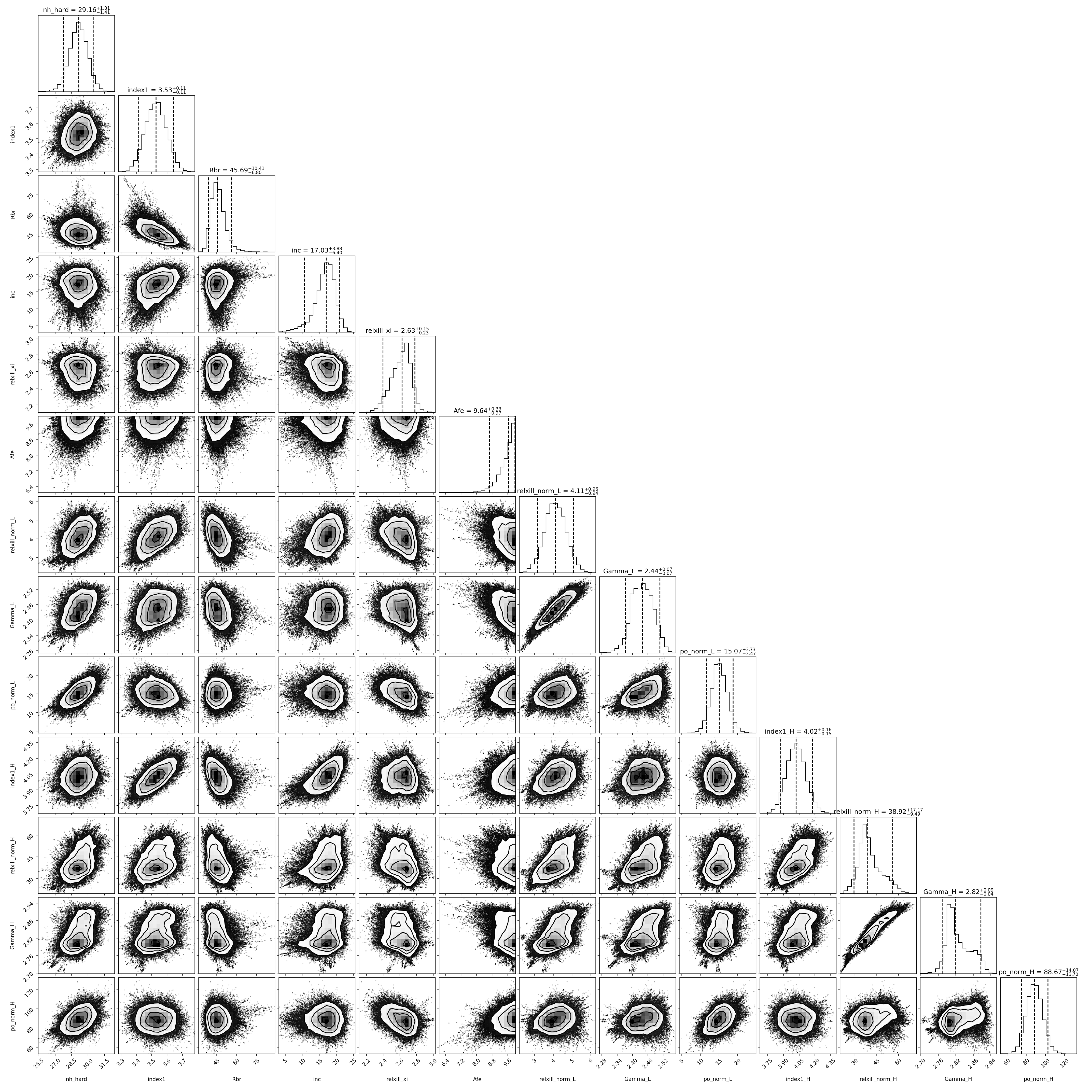}
	\caption{The corner plots for the MCMC best fit for the hard band components.Parameters that lack the subscript L or H denote values that were linked between epochs, while parameters with the subscript L indicate the low state value and parameters with subscript H indicate the high state value. Note that nh\_hard is the column density of the absorber applied to the continuum components.}
	\label{fig:hardMCMC}
\end{figure*}
\begin{figure}
	\centering
	\includegraphics[width=0.5\linewidth]{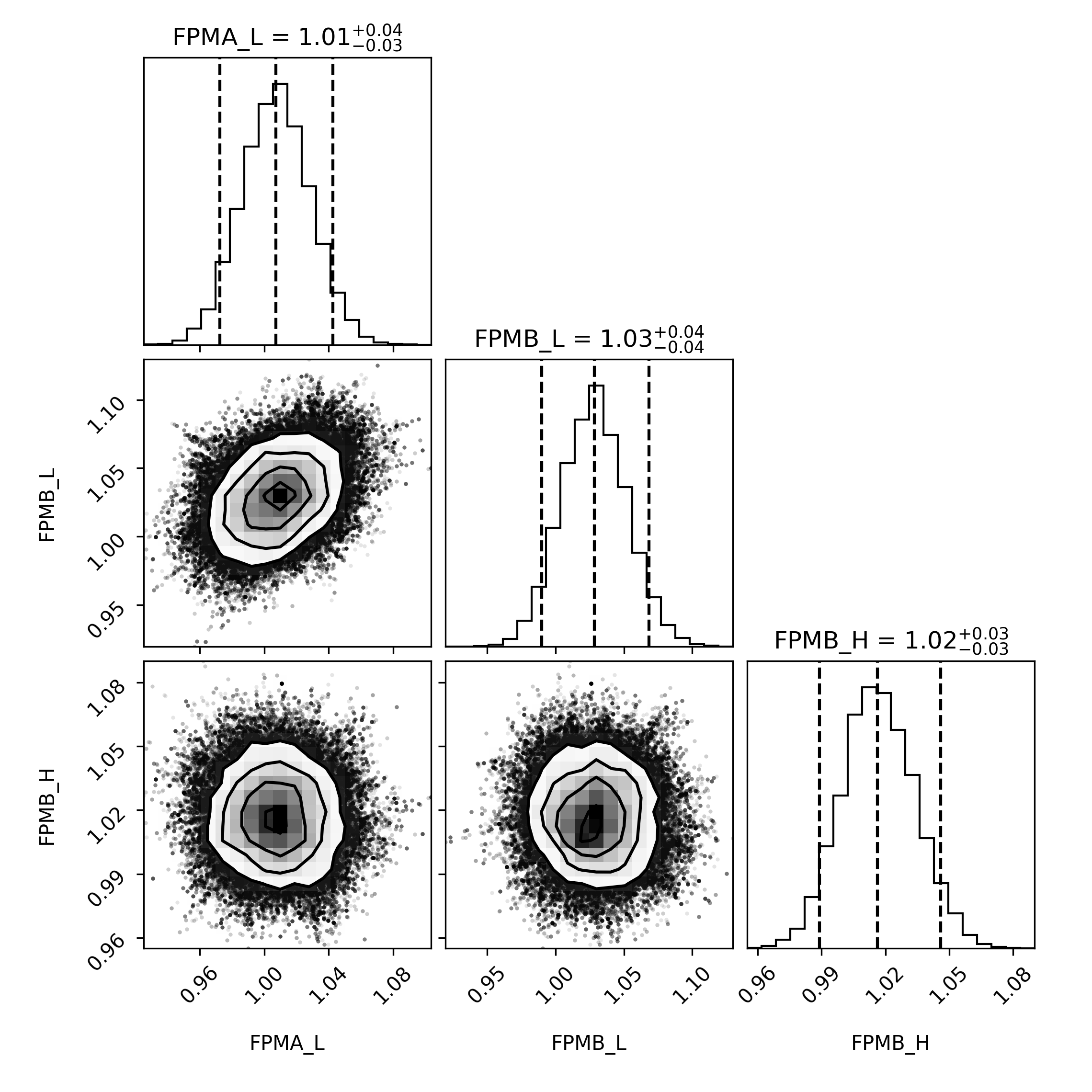}
	\caption{The corner plots for the MCMC best fit for the calibration constants. parameters with the subscript L indicate the low state calibration constants while the parameter with the subscript H indicates the high state calibration constant.}
	\label{fig:calMCMC}
\end{figure}
\end{document}